\documentclass[usenatbib, onecolumn]{mn2e}
\usepackage{graphicx}

\begin{document}
\title[Gravitational Balescu-Lenard equation]{A Balescu-Lenard type kinetic equation for the\\
collisional evolution of stable self-gravitating systems}

\author[J. Heyvaerts]{J. Heyvaerts$$\\
$$ Observatoire Astronomique de Strasbourg, \\
11, rue de l'Universit\'e, 67000 Strasbourg, France}

\maketitle
\begin{abstract}
A kinetic equation for the collisional evolution of stable,
bound, self gravitating and slowly relaxing systems is established, which is valid when 
the number of constituents is very large. 
It accounts for the detailed dynamics and 
self consistent dressing by
collective gravitational interaction of the colliding particles, for the system's
inhomogeneity and for different constituent's masses. It describes the coupled evolution 
of collisionally interacting populations, such as stars in a thick disk
and the molecular clouds off which they scatter.\\
The kinetic equation derives from the BBGKY hierarchy in the limit of 
weak, but non-vanishing,  
binary correlations, an approximation which is well justified for 
large stellar  systems. 
The evolution of the one-body distribution function is described in action angle space. The
collective response is calculated using a biorthogonal basis of pairs of density-potential functions.\\
The collision operators are expressed in terms of the collective response function 
allowed by the existing distribution functions at any
given time and involve particles in resonant motion. These equations are shown to satisfy an H-theorem.
Because of the inhomogeneous character of the system, the relaxation causes the
potential as well as the orbits of the particles to secularly evolve. 
The changing orbits also cause the angle Fourier coefficients of the basis potentials to change with time.
We derive the set of equations which describes this coupled evolution 
of distribution functions, potential and basis Fourier coefficients
for spherically symmetric systems.
In the homogeneous limit, which sacrifices  the description of the evolution of 
the spatial structure of the system but retains the effect of collective gravitational dressing,
the kinetic equation reduces to a form similar to the Balescu-Lenard equation of plasma physics.
\end{abstract}

\section{Introduction and motivation}
\label{introduction}

The description of collisional relaxation in a self-gravitating system usually rests on
a Fokker-Planck equation in which the diffusion and braking coefficients are
calculated in the local approximation, taking the finite dimension of
the system into account by limiting the impact parameter of the collisions to a length of order
of the system's size \citep{Chandra, Chandra43, BinneyTremaine, SpitzerAmas}.
Although characteristic relaxation
times may be somewhat overestimated by this approximation due to the neglect 
of collective self-gravitational effects \citep{Weinberg93},
such a kinetic equation may provide in practice a reasonable description
of the collisional relaxation of gravitationally bound systems.
It nevertheless rests on assumptions which, from a principle point of view,  are unsatisfactory  because
the motion of particles during the collision is regarded as rectilinear and uniform
and the system's inhomogeneity, which is basically the reason why 
collisions with an infinite impact parameter do not occur, is treated by way of an ill-defined cutoff. 
Moreover, the collective response of the system is not taken into account, since
the Fokker-Planck collision term only considers binary collisions between naked particles.
A self-gravitating medium, unlike an electrical plasma, does not respond 
to the presence in it of a particle by screening 
its interaction potential with other particles.
As a result, even distant particles effectively interact, while in electrical, globally neutral, 
plasmas, the effective interaction distance is limited to the Debye length.
In a self-gravitating system, the distance between interacting particles
is only limited by the system's inhomogeneity.  
The spatial structure of the system matters as well as the details of the particle orbits.

\bigskip

The consistent inclusion 
of collective screening effects in a kinetic equation for electrically interacting weakly coupled particles 
has been one of the major theoretical achievements in plasma physics when 
\citet{Balescu} and \citet{Lenard} could derive an equation surpassing 
in consistency the simple Fokker-Planck equation \citep{petitSpitzer}.
It is the aim of this paper to derive  a similar equation for
self-gravitating systems. The task is slightly more difficult because the screening 
of the electrical interaction at the, usually small, Debye length allows,  in electrically interacting systems,
to take the homogeneous and uniform motion limits. These limits cannot be taken 
in a self-gravitating system. 
We overcome this difficulty by expressing the kinetic equation in 
action angle space rather than in position momentum space. This is possible when the Hamiltonian
corresponding to the average potential $U({\mathbf{r}})$ of the system is integrable.
It is nevertheless uneasy in general
to toggle from one to the other space, although this is certainly possible for spherically
symmetric potentials, for flat systems (which may however be unstable) and for special
thick disk potentials. Numerical methods could be used to achieve the necessary transformation
\citep{PichonCannon, McMillanBinney}.
As an illustrative example, we shall give special attention to spherically symmetric potentials,
expanding their kinetic equation into 
a system which almost entirely avoids any calculation in the position-momentum space. 
The system's inhomogeneity requires that solutions to the Poisson equation are
easily found for any inhomogeneous mass distributions. This is achieved by projecting on a 
biorthogonal basis of pairs of density-potential functions.

\bigskip

Many astrophysical systems which have evolved to a quasi-stationary
collisionless equilibrium still keep evolving on time scales longer than the
dynamical time as a result of gravitational noise induced by their own constituents
or by external ones.
We disregard external perturbators, which we define as
unbound to the system, although, as did \citet{Weinberg2001II}, these could be treated,
if numerous and frequent enough, as a given, non-evolving, 
population providing a source of gravitational noise for other populations.
Loosely bound satellites or remote star populations are regarded as
internal to the system. This is possible because our set of kinetic equations allows to simultaneously
follow different mass populations. Dwarf satellite galaxies
could be regarded for example as one such mass population.
Globular clusters, dwarf galaxies, disk galaxies and their haloes
are examples of bound systems still evolving as a result of internal noise caused by particle discreteness.
Such systems are the object of our study. As in any weakly coupled system, the particles  suffering collisions
are dressed by the polarization clouds caused by their own influence on other particles. 
Collisions between dressed particles have quantitatively different outcomes                                             
than collisions between naked ones \citep{Weinberg98}. This may reflect
in significant differences in calculated effective relaxation times and braking or diffusion coefficients, especially when the
system, though stable, is not too far from instability \citep{Weinberg93}.
It is therefore useful to account for collective dressing when calculating such processes as secular thick disk evolution, 
mass segregation in galaxies or in star clusters, or the damping by dynamical friction
of galactic populations on high energy orbits.
For simplicity, the kinetic equations to be derived below assume that the system is
stationary on a dynamical time scale. They thus cannot address questions in which the distribution
in angle variable matters,
such as the dissolution of freshly accreted satellites,             
although a simple extension of the theory could. Since however our equations describe the coupled evolution of
all populations present in the system, they are well suited to
study, for example, the simultaneous
evolution by dynamical friction and diffusion of a stellar population 
and the population of molecular clouds off which these stars scatter.

\bigskip

The collective response of a self gravitating system to the presence of a perturbing body 
has been considered by a number of authors, analytically \citep{Weinberg89, Weinberg95,
MuraliTremaine, SahaJog} or numerically
\citep{ThielheimWolff, GnedinOstriker}. Sometimes, 
the reaction of this perturbation on the perturbing body itself is calculated,
as did \citet{Kalnajs72}, who computed the drag on a large body moving in an homogeneous medium,
taking the collective response of this medium into account, and 
\citet{WeinbergTremaine}, who considered the global, self-consistent, 
perturbation caused by a satellite or a barred structure in a
spherically symmetric system and its reaction on the perturbator object by the effect of dynamical friction.
The secular evolution of the system in response to such 
perturbations has been considered by \citet{Weinberg2001}, who considered general types
of perturbations on a galaxy, and by \citet{PichonAubert} who considered perturbations caused by the cosmological
environment on dark matter haloes. This evolution is of course
in principle observable in N-body simulations, which however have their own difficulties in calculating the long term
evolution of such systems \citep{Binney2004}. 
A number of authors \citep{Murali, Weinberg2001, PichonAubert} 
have studied the collective perturbations caused in a massive spherical galactic halo by its environment.
They could calculate the response of this system 
by resorting to a representation of the particle's motion in action and angle variables, a method first
used by \citet{Kalnajs77}. We follow them on this road.
They also made good use of a basis of biorthogonal pairs of density-potential functions.
\citet{Weinberg93} first derived a kinetic equation for the collisional 
relaxation of a self gravitating system along these lines. His equation accounts
for the self-consistent gravitational dressing of the particles, but is otherwise simplified, the
geometry supposedly being that of an homogeneously filled periodic cube.
The inhomogeneous nature of the system should be described more accurately, still accounting for 
collective gravitational dressing effects. This is specifically the aim of this paper. \citet{Chavanis}
presented a similar approach to ours for one-dimensional 
systems, the constituents of which interact by a general long range force. 
In this paper we further elaborate in section \ref{secsystemcouple} on the structural evolution of
the inhomogeneous system and on the secular evolution of the orbits.

\section{Cutting the BBGKY hierarchy}
\label{secfromBBGKYtoBL}

\subsection{Reduction of the hierarchy to a kinetic equation}
\label{subsecBBGKYtokinetic}

The Liouville equation for the N-body distribution function of a system of interacting particles can be
translated into a hierearchy of equations, the BBGKY hierarchy,
for the reduced 1-body, 2-body, 3-body etc .. distribution functions \citep{Balescubook, BinneyTremaine}. 
The equation for the 1-body distribution function also involves the 2-body distribution,
the equation for the 2-body distribution involves the 3-body distribution and so on.
The kinetic equation being meant to be an autonomous equation for the 1-body distribution 
$f_1(\mathbf{r}, \mathbf{p}, t)$, its derivation necessarily 
involves some approximation allowing to cut this hierarchy. This is  usually 
done at the level of the equation 
of evolution of the 2-body distribution function, reducing it to a relation between the
2-body and the 1-body distribution functions.
The simplified equation for the 2-body distribution 
$f_2(\mathbf{r}_1, \mathbf{p}_1, \mathbf{r}_2,  \mathbf{p}_2,t)$ is then solved
in terms of the 1-body distribution $f_1(\mathbf{r}_1, \mathbf{p}_1, t)$
and the result, once introduced in the first equation of the hierarchy, 
provides the desired kinetic equation for $f_1$.

\bigskip

Plasma physics knows of two such successful approximations:
rare and short range interactions, allowing to ignore 3-body collisional effects on the evolution
of the 2-body distribution function, leading
to the Boltzmann equation \citep{Uhlenbeck} and weakly coupled, collective, systems
in which the 3-body correlations may be neglected and the 2-body correlations considered weak, leading to
the Balescu-Lenard equation \citep{Balescu, Lenard}. 
The weak correlation approximation
is valid when the number of particles in the effective interaction sphere, the Debye sphere, 
is large. This approximation is also valid for self-gravitating systems  
with a large number $N$ of simultaneously interacting particles. The coupling in this case is indeed weak, 
the ratio of the average interaction
energy to the average kinetic energy scaling as $N^{-2/3}$.
This provides a solid basis for the derivation of a kinetic equation. The larger $N$, the 
more valid the approximation is. For systems with a very large number of bodies, the resulting kinetic equation 
is almost exact, but for the description of strong collisions. 

\bigskip

The constituents of the system are considered to be point-like objects of different masses, 
which we refer to as particles. They need not all be stars,
but could be other entities as well, such as molecular clouds, 
bound clusters, a population of satellites or lumps of dark matter in the halo of a galaxy. 
The kinetic equations to be derived below are valid as long as most collisions are weak,
which implies that the collisional evolution time of any type of particles remains long compared to
the dynamical time. We  assume that the masses of the consituents 
come in a finite set. Each mass group is labeled by a lower case latin letter.

\subsection{Notations}
\label{subsecnotations}

An efficient and concise notation is needed.
Some weakly relevant
variables, such as time, will often be omitted from the list
of arguments of some functions.
The subscripts 1 or 2 on one- or two-body distributions or correlation functions
will also be omitted, the number of arguments indicating the number of bodies involved.
The 1-body  distribution function of particles of species $a$ (that is, of mass $m_a$) 
is denoted by $f^{a}$, the 2-body  distribution function of a pair of particles 
of species $a$ and $b$ (where $a$ and $b$ may be equal or
different) is $f^{ab}$ and the corresponding   
2-body correlation function is $g^{ab} = f^{ab} - f^{a}f^{b}$. 
The space and momentum integral of a 1-body distribution function is the total number
of particles of the considered species. Similarly, the space and momentum integral of 
2-body distribution functions is the total number of pairs of the considered species. When $a =b$, 
pairs should be regarded as ordered entities.
The position and momentum $(\mathbf{r}_1, \mathbf{p}_1)$ of a particle 
is simply noted 1, for brevity.
The three angle and three action variables of this  
particle similarly form a pair of vectors $(\mathbf{w}_1, \mathbf{J}_1)$. The same shorthand notation, 1,
is used where the context commands. 
The notation $d1$ represents either $d^3r_1 d^3p_1$ or $d^3w_1 d^3J_1$. These phase space volume
elements are equal because both sets of variables are canonical. 
The velocity of particle 1 is ${\mathbf{v}}_1$. 
The gradient with respect to a vectorial variable $\mathbf{u}$, like $\mathbf{r}$, $\mathbf{p}$,
$\mathbf{w}$ or $\mathbf{J}$, is denoted by ${\mathbf{\nabla}}_{\mathbf{u}}$. 
The derivative with respect to time is noted $\partial_t$. 

\bigskip

$G$ being Newton's constant, the gravitational force suffered by 
a particle of species $a$ with dynamical variables 1
(that is at $\mathbf{r}_1$ with momentum $\mathbf{p}_1$)
from a particle of species $b$ with dynamical variables 2 is:
\begin{equation}
{\mathbf{F}}_{ab}(1,2) = G m_a m_b \ \, \frac{\mathbf{r}_2 - \mathbf{r}_1}{\mid \! \mathbf{r}_2 - \mathbf{r}_1\!\mid^3}\ \cdot
\label{Fbadef}
\end{equation}
We ignore any external force, be it tidal or  exerted by some closeby external body.
The collective gravitational force ${\mathbf{F}}^0_{a}(1)$ exerted at $\mathbf{r}_1$ 
on a particle of species $a$ is the 1-body and species average of ${\mathbf{F}}_{ab}(1,2)$:
\begin{equation}
{\mathbf{F}}^0_{a}(1) = \sum_b \int\!d2 \ {\mathbf{F}}_{ab}(1,2) \, f^{b}(2) \ \cdot
\label{forcecolldef}
\end{equation}
The gravitational potential $U({\mathbf{r}}_1)$ from which this force 
derives is:
\begin{equation}
U(\mathbf{r}_1) = - \sum_b \int\!d2  \ \frac{ G m_b}{\mid \! \mathbf{r}_2 - \mathbf{r}_1\!\mid} \, f^{b}(2)\ \cdot
\label{potcollectifdef}
\end{equation}

\subsection{Weak correlations in terms of one-body distributions}
\label{gdepropagateur}

The first equation of the BBGKY hierarchy can be written:
\begin{equation}
\partial_t f^{a}(1) +  {\mathbf{v}}_1 \cdot {\mathbf{\nabla}}_{\mathbf{r}_1} f^{a}(1) + {\mathbf{F}}^0_{a}(1) 
\cdot {\mathbf{\nabla}}_{\mathbf{p}_1} f^{a}(1) =
-  \sum_b \int\!\! d2 \ {\mathbf{F}}_{ab}(1,2) \cdot {\mathbf{\nabla}}_{\mathbf{p}_1} g^{ab} (1,2) \ \cdot
\label{BBGKY1}
\end{equation}
Neglecting 3-body correlations, 
the second equation of the BBGKY hierarchy can be written as:
\begin{eqnarray}
&&\partial_t g^{ab} (1,2) + \Big({\mathbf{v}}_1 \!\cdot\! {\mathbf{\nabla}}_{\mathbf{r}_1} 
+ {\mathbf{v}}_2\! \cdot\! {\mathbf{\nabla}}_{\mathbf{r}_2}\Big) \,  g^{ab} (1,2)
+ \Big({\mathbf{F}}^0_{a}(1) \!\cdot\! {\mathbf{\nabla}}_{\mathbf{p}_1} 
+ {\mathbf{F}}^0_{b}(2) \!\cdot\! {\mathbf{\nabla}}_{\mathbf{p}_2}\Big)  \, g^{ab} (1,2) \ 
\nonumber \\
&& + \!  \sum_c \int\!d3 \ \, g^{bc}(2,3)\ {\mathbf{F}}_{ac}(1,3)  \!\cdot\! {\mathbf{\nabla}}_{\mathbf{p}_1} f^{a}(1) 
+ \!  \sum_c \int\!d3 \ \, g^{ac}(1,3)\ {\mathbf{F}}_{bc}(2,3)  \!\cdot\! {\mathbf{\nabla}}_{\mathbf{p}_2} f^{b}(2) 
\! = {\mathbf{F}}_{ab}(1,2) \cdot ({\mathbf{\nabla}}_{\mathbf{p}_2} - {\mathbf{\nabla}}_{\mathbf{p}_1}) f^{a}(1) f^{b}(2)\cdot
\label{BBGKY2}
\end{eqnarray}
Equation (\ref{BBGKY2}) is linear in the correlation function and has on its right hand side
a source term $S^{ab}(1,2,t)$ which is a functional of the 1-body distribution functions, namely:
\begin{equation}
S^{ab}(1,2,t) =
{\mathbf{F}}_{ab}(1,2) \cdot ({\mathbf{\nabla}}_{\mathbf{p}_2} - {\mathbf{\nabla}}_{\mathbf{p}_1}) f^{a}(1) f^{b}(2) \ \cdot
\label{sourceterm}
\end{equation}
The solution for
$g^{ab} (1,2,t)$ can be found in terms of the sources $S$ by working out the Green's function,
or propagator, of the operator on the left hand side of equation (\ref{BBGKY2}).
This Green's function is a matrix in particle species space, ${\cal{G}}^{ab}_{pq}(1,2,1',2',\tau)$, 
in terms of which the correlation function can be expressed as:
\begin{equation}
g^{ab} (1,2,t) = \sum_{p,q} \int_0^{\infty}\!\!\!  d\tau\!\! \int \! d1'\!\! \int\! d2'
\ \ {\cal{G}}^{ab}_{pq}(1, 2, 1', 2', \tau)\,  S^{pq}(1', 2', t - \tau) \ \cdot
\label{Green2}
\end{equation}
Equation (\ref{Green2}) expresses the correlation function  as a functional $g^{ab}(1,2;f)$ of the 1-body distributions.
Once the 2-body propagator has been found, the solution (\ref{Green2}) for $g^{ab}(1,2)$
may be substituted on the right hand side of equation (\ref{BBGKY1}), which then depends explicitly, and only,
on the 1-body distributions. We call it
the collision operator ${\cal{C}}^a(f)$ for species $a$:
\begin{equation}
{\cal{C}}^a(f) = -  \sum_b \int\!\! d2 \ \  {\mathbf{F}}_{ab}(1,2) \cdot {\mathbf{\nabla}}_{\mathbf{p}_1} g^{ab} (1,2; f)\ \cdot
\label{poseCa}
\end{equation}
The initial value of the 2-body propagator is: 
\begin{equation}
{\cal{G}}^{ab}_{pq}(1,2,1',2',0)= \delta^a_p\, \delta^b_q\, \delta(1 -1')\, \delta(2-2') \ ,
\label{Greeninitiale}
\end{equation}
where $\delta(1 - 1')$ is a Dirac function and $\delta^a_p$ a Kronecker symbol.
By substituting equation (\ref{Green2}) in equation (\ref{BBGKY2}), 
it can be shown that the  2-body
propagator can be factored into the product of two 1-body propagators:
\begin{equation}
{\cal{G}}^{ab}_{pq}(1,2, 1', 2',\tau)  = {\cal{G}}^{a}_p(1, 1', \tau)\,  {\cal{G}}^{b}_q(2, 2', \tau) \ \cdot
\label{factorpropagateur}
\end{equation}
Had we considered strong interactions as well, the  correlation function $g^{ab}(1,2)$ would not have been negligible
compared to $f^a(1) f^b(2)$ and the right hand side term of equation (\ref{BBGKY2}) would have been
changed by the substitution of $f^a\!f^b\!+\!g^{ab}$ to $f^a\!f^b$. 
The 2-body propagators would in this case not factor as in equation (\ref{factorpropagateur}).
In the weak correlation approximation considered here,
the 1-body propagators ${\cal{G}}^{a}_p(1, 1',\tau)$ satisfy the linearized Vlasov equations:
\begin{equation}
\partial_\tau \, {\cal{G}}^{a}_p(1, 1',\tau) + 
({\mathbf{v}}_1 \!\cdot\! {\mathbf{\nabla}}_{\mathbf{r}_1} + {\mathbf{F}}^0_{a}(1)\! \cdot \! {\mathbf{\nabla}}_{\mathbf{p}_1}) 
\, {\cal{G}}^{a}_p(1, 1',\tau) 
 +  \left(\sum_c \int d2 \, {\cal{G}}^{c}_p(2, 1',\tau)\  {\mathbf{F}}_{ac}(1,2)\right) \cdot 
{\mathbf{\nabla}}_{\mathbf{p}_1} f^a(1) = 0 \ ,
\label{VlasovlinpourG1}
\end{equation}
with initial condition ${\cal{G}}^{a}_p(1, 1',0) = \delta^a_p \delta(1-1')$.
The solution of equation (\ref{VlasovlinpourG1})
has to be found for $\tau \ge 0$ only, because of causality.
According to Bogoliubov's synchronisation hypothesis \citep{Bogoliubovsynchro}, 
the 1-body distribution functions  
can be regarded as constant in equations (\ref{Green2}) and
(\ref{VlasovlinpourG1}) because they evolve on the relaxation time scale,
which is much longer than the time 
required for the correlation function to reach an equilibrium, given the present value of the 
1-body distributions. 
The correlations at a given time $t$ then are functionals 
of the one particle distribution functions at this very same time. 

\bigskip

Equation (\ref{VlasovlinpourG1}) can be solved by means of a Laplace transform 
with respect to the time lapse $\tau$.
The Laplace transform $f(\omega)$ of a function of time $f(t)$ depends on a complex argument $\omega$. 
The transformation and its inverse are defined by:
\begin{equation}
f(\omega) = \int_0^\infty \! f(t)\, e^{i\omega t} dt \qquad {\mathrm{and}} \qquad 
f(t) = \frac{1}{2 \pi} \int_B  f(\omega) \, e^{-i\omega t} d\omega \ \cdot
\end{equation}
The direct transform is convergent only when the imaginary part of $\omega$ exceeds some ordinate
of convergence, above which the function $f(\omega)$ is regular. Below it, it is defined by analytical continuation.
The Bromwich contour $B$ which appears in the inverse  transformation runs parallel to the real axis from 
$-\infty$ to $+\infty$ above all singularities of $f(\omega)$.  
Equation (\ref{VlasovlinpourG1}) is Laplace-transformed into:
\begin{equation}
-i\omega \, {\cal{G}}^{a}_p(1, 1',\omega) 
+ ({\mathbf{v}}_1 \!\cdot\! {\mathbf{\nabla}}_{\mathbf{r}_1} \,
+ \, {\mathbf{F}}^0_{a}(1)\! \cdot\, {\mathbf{\nabla}}_{\mathbf{p}_1}) 
 \ {\cal{G}}^{a}_p(1, 1',\omega)
+ \sum_c \int\! d2 \ {\cal{G}}^{c}_p(2, 1',\omega) \ \,  {\mathbf{F}}_{ac}(1,2) \cdot
{\mathbf{\nabla}}_{\mathbf{p}_1} f^a(1) \, =  \, \delta^a_p \, \delta(\!1 -1') \ \cdot
\label{VlasovlinpourG1Lapl}
\end{equation}

\section{Particle motions and basis functions in angle and action variables}
\label{secangleaction}

\subsection{Angle and action variables}
\label{subsecangleaction}

The particle motions in the self gravitational field are complex in general. This
precludes a direct solution of equation (\ref{VlasovlinpourG1}) by integration along unperturbed trajectories. 
It is preferable to change the position and momentum variables for a set of canonical 
angle and action variables \citep{Goldstein}. So doing, the description of the motion becomes simple, all the
complexity being embodied in the relation between position and momentum variables and
angle and action variables. 
By definition, the  
Hamiltonian $\cal{H}$ in angle and action variables 
depends only on the three actions $J_1$, $J_2$, $J_3$, which we regard as the three components
of an action vector ${\mathbf{J}}$. The three actions are constants
of the motion and the three angles $w_1, w_2, w_3$ which similarly  form
the components of an angle vector ${\mathbf{w}}$, 
vary linearly in time. The angular frequency of the angle $w_i$ is $\Omega_i = {\partial {\cal{H}}}/{\partial J_i}$.
The frequencies $\Omega_i$ form the components of a frequency vector ${\mathbf{\Omega}}$ which
depends on ${\mathbf{J}}$. For brevity, we use shorthand notations, such as:
\begin{equation}
{\mathbf{\Omega}}_1 \equiv {\mathbf{\Omega}}({\mathbf{J}}_1)\ ,  \qquad \qquad 
{\mathbf{\Omega}}'_1 \equiv {\mathbf{\Omega}}({\mathbf{J}}'_1) \ \cdot
\label{Omegasreduits}
\end{equation}
The derivative following the motion is 
(${\mathbf{v}} \cdot {\mathbf{\nabla}}_{\mathbf{r}} + {\mathbf{F}}^0 \cdot {\mathbf{\nabla}}_{\mathbf{p}}$). 
The actions being first integrals, this operator
translates in angle and action variables into $(d{\mathbf{w}}/dt) \cdot {\mathbf{\nabla}}_{\mathbf{w}}$, that is,
$
{\mathbf{v}} \cdot {\mathbf{\nabla}}_{\mathbf{r}} + {\mathbf{F}}^0 \cdot {\mathbf{\nabla}}_{\mathbf{p}}
= {\mathbf{\Omega}} \cdot {\mathbf{\nabla}}_{\mathbf{w}}
$
The  last, collective, term of the left hand side of equation (\ref{VlasovlinpourG1Lapl}) must
be expressed in action and angle variables.
It is of the frequently met general form:
\begin{equation}
\lambda \, \int\!\! d2 \, M(2)\,  
\frac{{\mathbf{r}}_2 - {\mathbf{r}}_1}{\mid\! {\mathbf{r}}_2 - {\mathbf{r}}_1\!\mid^3}
\cdot  {\mathbf{\nabla}}_{\mathbf{p}_1} N(1) \ \cdot
\label{laformegenerale}
\end{equation}
The force in equation (\ref{laformegenerale}) can be expressed in terms of a "potential" $\phi$ such that:
\begin{equation}
\int\!\! d2 \, M(2)  \   
\frac{{\mathbf{r}}_2 - {\mathbf{r}}_1}{\mid\! {\mathbf{r}}_2 - {\mathbf{r}}_1\!\mid^3} =
-  \, {\mathbf{\nabla}}_{\mathbf{r}_1} \phi({\mathbf{r}_1})  \ \cdot
\label{introdephi}
\end{equation}
This potential and the "mass distribution" $D$ from which it derives, depend
on the function $M(2)$ only. They are defined by:
\begin{equation}
\phi ({\mathbf{r}_1}) = - \, \int \!\! d2 \ \frac{ M(2)}{\mid \! {\mathbf{r}_2} - {\mathbf{r}_1} \! \mid}
\ , \qquad \qquad \qquad 
D({\mathbf{r}_2}) = \int\!\! d^3\!p_2 \ M(2) \ \cdot
\label{associatedD}
\end{equation}
Since $\phi(1)$ is independent of ${\mathbf{p}_1}$,
${\mathbf{\nabla}}_{\mathbf{r}_1} \phi(1) \!\cdot\! {\mathbf{\nabla}}_{\mathbf{p}_1} N(1)$
is the Poisson bracket $\{\, \phi(1), N(1)\, \}$.
This bracket being invariant on a change of canonical variables,
the expression (\ref{laformegenerale}) can be written as:
\begin{equation}
\int\!\! d2 \, M(2)\,
\frac{{\mathbf{r}}_2 - {\mathbf{r}}_1}{\mid\! {\mathbf{r}}_2 - {\mathbf{r}}_1\!\mid^3}
\cdot  {\mathbf{\nabla}}_{\mathbf{p}_1} N(1) =
- {\mathbf{\nabla}}_{\mathbf{r}_1} \phi \, \cdot {\mathbf{\nabla}}_{\mathbf{p}_1} N(1) \, = \, - \Big(
{\mathbf{\nabla}}_{\mathbf{w}_1}\phi \cdot {\mathbf{\nabla}}_{\mathbf{J}_1} N(1)
- {\mathbf{\nabla}}_{\mathbf{J}_1} \phi \cdot {\mathbf{\nabla}}_{\mathbf{w}_1} N(1)\Big) \ \cdot
\label{PoissonbracketwJ}
\end{equation}
All functions depend periodically, with period $2\pi$,
on the angles, with respect to which a discrete Fourier transform can be made.
All components of the associated wave vector ${\mathbf{k}}$ are relative integers.
The transform of any function $f({\mathbf{w}}, {\mathbf{J}})$  and the inverse
transform are defined by:
\begin{equation}
f({\mathbf{w}}, {\mathbf{J}}) = \sum_{\mathbf{k}} f_{\mathbf{k}}({\mathbf{J}})  \,
e^{ i \, {\mathbf{k}}\cdot {\mathbf{w}} }
\qquad {\mathrm{and}} \qquad
f_{\mathbf{k}}({\mathbf{J}}) = \! \int\!\!\!\! \int\!\!\!\!\int \frac{d^3w}{8\pi^3} \, f({\mathbf{w}}, {\mathbf{J}})
\, e^{ - i \, {\mathbf{k}}\cdot {\mathbf{w}} } \ \cdot
\label{defexpansFourierw}
\end{equation}
Each integral in the second term of equation (\ref{defexpansFourierw})
is over the $2\pi$ period of one of the components of ${\mathbf{w}}$.
The transform of the Dirac function $\delta({\mathbf{w}})$ is $1/8\pi^3$ and the
transform of unity is $\delta(\mathbf{k})$, where $\delta$ is here a triple Kronecker symbol.
The position ${\mathbf{r}}$ of a particle is a function of its angle and action variables, ${\mathbf{w}}$ and ${\mathbf{J}}$.
The simple Fourier transforms with respect to the
angles ${\mathbf{w}}_1$ of $\psi^\alpha({\mathbf{r}}_1)$ and ${\cal{G}}^a_p(1, 1', \omega)$
and the double Fourier transform of the propagator with respect
to angles ${\mathbf{w}}_1$ and ${\mathbf{w}}'_1$ are:
\begin{equation}
\psi^\alpha(1) \ \leftrightarrow \ \psi^\alpha_{\mathbf{k}_1}({\mathbf{J}}_1)
\qquad \qquad \qquad
{\cal{G}}^a_p(1, 1', \omega) \ \leftrightarrow \ G^{ap}_{\mathbf{k}_1} ({\mathbf{J}}_1, 1', \omega)
\qquad \qquad \qquad
{\cal{G}}^a_p(1, 1', \omega) \ \leftrightarrow \ G^{ap}_{ {\mathbf{k}}_1 {\mathbf{k}}'_1 }
({\mathbf{J}}_1, {\mathbf{J}}'_1, \omega) \ \cdot
\end{equation}

\subsection{Biorthogonal density-potential bases}
\label{subsecbiorthbases}

A basis of biorthogonal density-potential pairs
is effective in calculating the potential $\phi(1)$ defined by eq. (\ref{associatedD}).  
Many such bases have been proposed \citep{Kalnajs71, Cluttonflat, Clutton3D, Kalnajs76,
AokiIye, Aokietal, Saha, 
HernquistOstriker, RobijnEarn, BrownPapa, RhamatiJallali}.
A basis element is labeled by a greek letter. The dummy index rule is used for these
basis indices. Let $D^\alpha({\mathbf{r}})$ and 
$\psi^\alpha({\mathbf{r}})$ be the density and the potential of the element $\alpha$ of the basis.
The potential $\psi^\alpha$ derives from the density distribution $D^\alpha$ and is related to it by:
\begin{equation}
\psi^\alpha({\mathbf{r}}) = - \, 
\int\!\!d^3\!r' \  \frac{D^\alpha({\mathbf{r}}')}{\mid\! {\mathbf{r}}' - {\mathbf{r}}\mid} \ \cdot
\label{DeltaetPsi}
\end{equation}
The basis is biorthogonal and normalized, such that:
\begin{equation}
 \int\!\! d^3\!r \ D^\alpha({\mathbf{r}}) \ (\psi^{\beta}({\mathbf{r}}))^* = -\ \delta^\alpha_\beta \ \cdot
\label{biortho}
\end{equation}
The symbol on the right of equation (\ref{biortho}) is a generalized Kronecker.
The minus sign results from the fact that when $\alpha = \beta$ 
the left hand side of equation (\ref{biortho}) necessarily is negative. 
The functions to be expanded on the basis being real,
the complex conjugates  of $D^\alpha$ and $\psi^\alpha$,
\begin{equation}
D^{\hat{\alpha}}({\mathbf{r}}) \!\equiv\! (D^{\alpha}({\mathbf{r}}))^*  \quad {\mathrm{and}} \quad
\psi^{\hat{\alpha}}({\mathbf{r}}) \!\equiv\!  (\psi^{\alpha}({\mathbf{r}}))^* \ ,
\label{hatetstar}
\end{equation}
also form an element $\hat{\alpha}$ of the basis, which in general is different from $\alpha$.
The variable ${\mathbf{r}}$ being a length and the Kronecker $\delta$
in (\ref{biortho}) being dimensionless, equations
(\ref{DeltaetPsi}) and (\ref{biortho}) imply that $D^\alpha$ and $\psi^\alpha$
have dimensions $L^{-5/2}$ and $L^{-1/2}$ respectively.
Any density distribution $D({\mathbf{r}})$ and its associated potential $\phi({\mathbf{r}})$
can be expanded on the basis as:
\begin{equation}
D({\mathbf{r}}) =  a_\alpha \, D^\alpha({\mathbf{r}})
\qquad \qquad 
\phi({\mathbf{r}}) = a_\alpha \psi^\alpha({\mathbf{r}}) \ \cdot
\label{phiexpansiongene}
\end{equation}
The basis functions $\psi^\alpha({\mathbf{r}})$
are not real in general, which implies that
$\psi^\alpha_{-{\mathbf{k}}}\! \neq \! (\psi^{\alpha}_{\mathbf{k}})^*$.
The notation $\psi^{\alpha *}_{\mathbf{k}}$ denotes the complex conjugate of
$\psi^{\alpha}_{\mathbf{k}}$. The notation $\psi^{\hat{\alpha}}_{\mathbf{k}}$
is adopted for the ${\mathbf{k}}$-Fourier transform
of the function $ \psi^{\hat{\alpha}}({\mathbf{r}}) \equiv  (\psi^{\alpha}({\mathbf{r}}))^*$.
In general, $\psi^{\hat{\alpha}}_{\mathbf{k}} \neq \psi^{\alpha*}_{\mathbf{k}}$.
Complex conjugation implies however that:
\begin{equation}
\psi^{\hat{\alpha}}_{\mathbf{k}} = \left(\psi^{\alpha}_{-\mathbf{k}}\right)^* \ \cdot
\label{psipsichapeau}
\end{equation}
The coefficients  of the expansions (\ref{phiexpansiongene}) can be calculated by using the 
biorthogonality relation (\ref{biortho}) and expressed in angle and action variables
by using the density-potential basis and angle Fourier coefficients.
In particular, the coefficient $a_\alpha$ associated with the density field of equation (\ref{associatedD}) is:
\begin{equation}
a_\alpha  =  - \int\!\!d2 \  M(2) \,  (\psi^{\alpha}(2))^* 
= - 8\pi^3 \sum_{\mathbf{k}} \int\!\!d^3\!J_2 \
\, M_{\mathbf{k}}({\mathbf{J}}_2)
\  \psi^{\alpha*}_{\mathbf{k}}({\mathbf{J}}_2) \ \cdot
\label{coeffadephib}
\end{equation}
The expression (\ref{laformegenerale}) is transformed in angle and action variables
by using the density-potential basis and angle Fourier coefficients into:
\begin{equation}
\lambda \, \int\! d2 \, M(2)  \ 
\frac{{\mathbf{r}}_2 - {\mathbf{r}}_1}{\mid\! {\mathbf{r}}_2 - {\mathbf{r}}_1\!\mid^3}
\cdot  {\mathbf{\nabla}}_{\mathbf{p}_1} N(1) \, = \, 
-\,\lambda \,  \sum_{{\mathbf{k}}_1} \ a_\alpha \, e^{i {\mathbf{k}}_1\cdot {\mathbf{w}}_1}
\Big( \psi^{\alpha}_{{\mathbf{k}}_1}(1) \, i {\mathbf{k}}_1 \! \cdot \!
{\mathbf{\nabla}}_{{\mathbf{J}}_1} N(1) - ( {\mathbf{\nabla}}_{{\mathbf{J}}_1} \psi^{\alpha}_{{\mathbf{k}}_1}(1)) 
\! \cdot \! {\mathbf{\nabla}}_{{\mathbf{w}}_1} N(1)\Big) \ ,
\label{expression14quasifinale}
\end{equation}
where the expansion coefficients $a_\alpha$ are given in terms of the function $M(2)$ by
equation (\ref{coeffadephib}). 
Other expressions of $a_\alpha$ will be established below for the case 
when $M(2)$ is the propagator,  as in equation (\ref{VlasovlinpourG1Lapl}).

\section{The linearized Vlasov propagator}
\label{secpropagator}

In a relaxing system, a collisionless equilibrium is supposedly reached on a
time scale shorter than the relaxation time, so that the system is stationary on the dynamical time scale.
This means that the distributions $f^a(1)$ really are functions $f^a({\mathbf{J}}_1)$ of the actions only.
The third, "collective",  term on the left of equation (\ref{VlasovlinpourG1Lapl}) is
of the form displayed in equation (\ref{laformegenerale}). The corresponding 
factor $\lambda$ and functions 
$N(1)$ and $M(2)$ particularize in this case to 
$\lambda = G m_a m_c$, $N(1) = f^a(1)$ and $M(2) = {\cal{G}}^c_p(2, 1',\omega)$. 
For these functions $N(1)$ and $M(2)$, the coefficients $a_\alpha $ of equation (\ref{coeffadephib}) are:
\begin{equation}
a^{cp}_\alpha(1',\omega) =  - 8\pi^3 \sum_{\mathbf{k}} \int\!\!d^3\!J \
\, G^{cp}_{\mathbf{k}}({\mathbf{J}}, 1', \omega)                                                
\  \psi^{\alpha*}_{\mathbf{k}}({\mathbf{J}}) \ \cdot
\label{adeFtransformG}
\end{equation}
Species-cumulative coefficients $A^p_\alpha$ are defined by:
\begin{equation}
A^p_\alpha(1', \omega) =  \sum_a \, m_a \, a^{ap}_\alpha (1', \omega) \ \cdot
\label{defA}
\end{equation}
Equation (\ref{VlasovlinpourG1Lapl}) for the 1-body propagators is
Fourier transformed with respect to ${\mathbf{w}}_1$ using equation 
(\ref{expression14quasifinale}), which gives:
\begin{equation}
G^{ap}_{\mathbf{k}_1}({\mathbf{J}}_1,1',\omega) = \frac{i}{8 \pi^3} \, \delta^a_p \, \delta({\mathbf{J}}_1 - {\mathbf{J}}'_1)
\, \frac{e^{-i\, {\mathbf{k}}_1\cdot {\mathbf{w}}'_1} }{ \omega - {\mathbf{k}}_1\cdot {\mathbf{\Omega}}_1 } - 
\ G m_a\ \frac{ \left({\mathbf{k}}_1\! \cdot\! {\mathbf{\nabla}}_{\mathbf{J}_1} f^a(1)\right) }{ 
\omega - {\mathbf{k}}_1 \cdot {\mathbf{\Omega}}_1 } \, 
\ \psi^\beta_{\mathbf{k}_1} ({\mathbf{J}}_1) \  A^p_\beta(1', \omega) \ \cdot
\label{eqGdekJnumero1}
\end{equation}
The coefficients $A$ (eq. (\ref{defA})) can be expressed in terms
of the Fourier transform of the propagators by using equation (\ref{adeFtransformG}):
\begin{equation}
A^p_\alpha(1',\omega) = - 8\pi^3 \sum_c \sum_{\mathbf{k}}\! \int\!\! d^3\!J \ \, m_c\, 
\psi^{\alpha*}_{\mathbf{k}}({\mathbf{J}}) \, G^{cp}_{\mathbf{k}}({\mathbf{J}}, 1', \omega) \ \cdot
\label{AetpropagateurG}
\end{equation}
Operating on equation (\ref{eqGdekJnumero1}) as on the function $G$ 
in equation (\ref{AetpropagateurG}), a linear system is obtained 
for the species-cumulative coefficients $A$. It can be written:
\begin{eqnarray}
&& \varepsilon^{\alpha \beta}(\omega) \, A^p_\beta(1', \omega) = \sigma^p_\alpha(1', \omega) \ ,
\label{eqpourAavecepsilon}
\\
&& \sigma^p_\alpha(1', \omega) = -i \, m_p \, \sum_{{\mathbf{k}}_1} \,
\frac{\psi^{\alpha*}_{\mathbf{k}_1}({\mathbf{J}}'_1) \ e^{-i \, {\mathbf{k}}_1 \cdot {\mathbf{w}}'_1 }  }{
\omega - {\mathbf{k}}_1 \! \cdot\!  {\mathbf{\Omega}}'_1 } \ ,
\label{sigma}
\\
&& \varepsilon^{\alpha \beta}(\omega) = \delta^{\alpha \beta}  - 
\sum_{a} \sum_{{\mathbf{k}}_1} \! \int\!\!d^3J_1 \ \,
8\pi^3 G m_a^2 \ \
\psi^{\alpha*}_{\mathbf{k}_1}(1)\,  \psi^{\beta}_{\mathbf{k}_1}\!(1)
\ \frac{{\mathbf{k}}_1\! \cdot\!  {\mathbf{\nabla}}_{\mathbf{J}_1} f^a(1) }{
\omega - {\mathbf{k}}_1 \! \cdot \! {\mathbf{\Omega}}_1 } \ \cdot
\label{epsilonalphabeta}
\end{eqnarray}
The solution of equation (\ref{eqpourAavecepsilon}),  
obtained by inverting the matrix $\varepsilon^{\alpha \beta}(\omega)$, 
is then introduced in equation (\ref{eqGdekJnumero1}), giving
the Fourier and Laplace transform of the propagator. So doing, 
a function ${\cal{D}}$ appears in the solution, which is defined by:
\begin{equation}
\frac{1}{{\cal{D}}_{{\mathbf{k}_1} {\mathbf{k}'_1}}({\mathbf{J}}_1, {\mathbf{J}}'_1, \omega)}
= \psi^\alpha_{\mathbf{k}_1}({\mathbf{J}}_1) \left(\varepsilon^{-1}(\omega)\right)^{\alpha \beta}
\psi^{\beta*}_{\mathbf{k}'_1}({\mathbf{J}}'_1) \ \cdot
\label{definitionD}
\end{equation}
Performing the inverse Fourier and Laplace transforms of equation (\ref{eqGdekJnumero1}),
the  1-body propagator itself is eventually found:
\begin{equation}
{\cal{G}}^{a}_{p}(1, 1', \tau) =  
\int_B \frac{d\omega}{2\pi} e^{-i \, \omega \tau}
\ \sum_{ {\mathbf{k}}_1} \sum_{ {\mathbf{k}}'_1} \ 
\frac{i\, e^{i ({\mathbf{k}}_1 \cdot {\mathbf{w}}_1- {\mathbf{k}}'_1\cdot {\mathbf{w}}'_1)} }{
8\pi^3(\omega - {\mathbf{k}}_1 \cdot {\mathbf{\Omega}}_1)  } 
\left(
\delta^a_p \, 
\delta({\mathbf{k}}_1\!\! -\!\! {\mathbf{k}}'_1) \, \delta ({\mathbf{J}}_1\! - \! {\mathbf{J}}'_1)  
+\,  
\frac{
8 \pi^3 G m_a m_p \left({\mathbf{k}}_1\!\cdot\! {\mathbf{\nabla}}_{\mathbf{J}_1} f^a(1)\right) 
}{
(\omega - {\mathbf{k}}'_1 \!\cdot\! {\mathbf{\Omega}}'_1) \, 
{\cal{D}}_{{\mathbf{k}_1} {\mathbf{k}'_1}}({\mathbf{J}}_1, {\mathbf{J}}'_1, \omega)   } 
\right) \ \cdot
\label{propagateur1body}
\end{equation}

\section{The kinetic equation}
\label{secEqLBgrav}

\subsection{Explicit writing of the kinetic equation}
\label{subsececrireLAequation}

The correlation function is obtained from the solution (\ref{propagateur1body}) for the 1-body propagator
by using equations (\ref{factorpropagateur}) and (\ref{Green2}). The kinetic equation and 
its collision operator are then given by equations (\ref{BBGKY1}) and (\ref{poseCa}).
Thanks to the Bogoliubov synchronisation hypothesis, this equation is local in time, because
the source term $S^{pq}(1',2', t-\tau)$ in equation (\ref{Green2}) can be regarded as independent of $\tau$ and  equal to its
value  at $\tau = 0$. The collision operator for the evolution of the distribution function of species $a$,
${\cal{C}}^a(f)$, is defined by equation (\ref{poseCa}) and can be written as:
\begin{eqnarray}
&&{\cal{C}}^a(f) = - \sum_{p,\, q} \int_0^\infty \!\!\! d\tau 
\int\! d1'\! \int\! d2' \! 
\int_B \! \frac{d\omega}{2\pi} \int_{B'} \! \frac{d\omega'}{2\pi}  e^{-i \, (\omega + \omega') \tau} \ 
\sum_b \int\! d2  \, \, \cdots
\nonumber \\
&&\qquad \qquad \cdots \,  
\left({\mathbf{F}}_{ab}(1,2) \cdot {\mathbf{\nabla}}_{{\mathbf{p}}_1}\right)
\ \, \Big( {\cal{G}}^a_p(1,1', \omega) \, {\cal{G}}^b_q(2,2',\omega') \, \ 
({\mathbf{F}}_{pq}(1',2') \cdot ({\mathbf{\nabla}}_{{\mathbf{p}}'_2} -
{\mathbf{\nabla}}_{{\mathbf{p}}'_1}))  \   f^p(1', t) f^q(2', t) \Big) \ \cdot
\label{Colloperateurbrut}
\end{eqnarray}
The somewhat lengthy transformations that must be performed 
to express this equation  in terms of the angle and action variables, 
of the density-potential basis and of its  angle Fourier transforms are 
described in appendix \ref{grossesmagouilles}. They eventually yield 
the following final form of the kinetic equations:
\begin{equation}
\partial_t f^{a}({\mathbf{J}}_1) 
= \sum_b \sum_{ {\mathbf{k}}_1 } \sum_{ {\mathbf{k}}_2 }\! \int\! d^3\! J_2
\ \, 8\pi^4 G^2 m_a^2 m_b^2 \  \
{\mathbf{k}}_1\! \cdot\! {\mathbf{\nabla}}_{ {\mathbf{J}}_1}
\ \Big(
\frac{\delta({\mathbf{k}}_1\!\cdot\!  {\mathbf{\Omega}}_1 - {\mathbf{k}}_2\!\cdot\!  {\mathbf{\Omega}}_2) }{
\left\vert
{\cal{D}}_{{\mathbf{k}}_1 {\mathbf{k}}_2}({\mathbf{J}}_1, {\mathbf{J}}_2, {\mathbf{k}}_1\!\cdot\!  {\mathbf{\Omega}}_1)
\right\vert^2  }
\ \left({\mathbf{k}}_1\! \cdot\! {\mathbf{\nabla}}_{ {\mathbf{J}}_1} \!
- \! {\mathbf{k}}_2\! \cdot\! {\mathbf{\nabla}}_{ {\mathbf{J}}_2 }\right) f^a({\mathbf{J}}_1)
f^b({\mathbf{J}}_2) \Big) \ , 
\label{LAequation}
\end{equation}
where $\cal{D}$ is defined by equation (\ref{definitionD}) and the response matrix 
elements $\varepsilon^{\alpha \beta}$ needed to determine $\cal{D}$ 
are expressed in terms of the 1-body distribution
functions by equation (\ref{epsilonalphabeta}). 
No convective term ${\mathbf{\Omega}}_1 \!\cdot\! {\mathbf{\nabla}}_{{\mathbf{w}}_1} f^a (1)$ appears
on the left hand side of eq.(\ref{LAequation}) because in a slowly relaxing system
the distribution functions $f^a(1)$ are meant to depend only on the actions. 

\bigskip

\subsection{Physical content of the kinetic equation}
\label{physicalcontent}

Equation (\ref{LAequation}) describes the relaxation of the distribution functions caused by the, supposedly weak, noise 
created by the discreteness of the particles accompanied by their associated 
gravitational polarization cloud \citep{Weinberg98, RostokerRosen}.
This is shown by working out the Fokker-Planck equation for 
the evolution of actions of the particles in this random field. The potential of a mass $m_2$ with action-angle variables
${\mathbf{J}}_2$, ${\mathbf{w}}_2$ on a particle $1$ with action-angle variables
${\mathbf{J}}_1$, ${\mathbf{w}}_1$ is:
\begin{equation}
{\tilde{U}}_2(1,t) = - \sum_{ {\mathbf{k}}_1 } \sum_{ {\mathbf{k}}_2 }\ G m_2
\ \frac{\exp(i ({\mathbf{k}}_1\!\cdot\!  {\mathbf{w}}_1 - {\mathbf{k}}_2\!\cdot\!  {\mathbf{w}}_2)) }{
{\cal{D}}_{{\mathbf{k}}_1 {\mathbf{k}}_2}({\mathbf{J}}_1, {\mathbf{J}}_2, {\mathbf{k}}_2 \!\cdot\!  {\mathbf{\Omega}}_2) } \ \cdot
\label{potuneparthabillee}
\end{equation}
The fluctuating part of the potential created by the discreteness of the dressed particles is the sum over all particles $2$ and all non-vanishing
${ {\mathbf{k}}_1 }$, ${ {\mathbf{k}}_2 }$ of potentials like (\ref{potuneparthabillee}). The rate of change 
of the action ${\mathbf{J}}_1$
of a particle $1$ in this fluctuating field is: 
\begin{equation}
{\dot{\mathbf{J}}}_1 = \sum_2 \sum_{ {\mathbf{k}}_1 \neq 0} \sum_{ {\mathbf{k}}_2 \neq 0} G m_1 m_2
\ i {\mathbf{k}}_1 \ \frac{\exp(i ({\mathbf{k}}_1\!\cdot\!  {\mathbf{w}}_1 
- {\mathbf{k}}_2\!\cdot\!  {\mathbf{w}}_2) }{
{\cal{D}}_{{\mathbf{k}}_1 {\mathbf{k}}_2}({\mathbf{J}}_1, 
{\mathbf{J}}_2, {\mathbf{k}}_2 \!\cdot\!  {\mathbf{\Omega}}_2) } \ \cdot
\label{evolJ1fluct}
\end{equation}
The braking and diffusion coefficients of the corresponding Fokker-Planck equation are obtained from,
respectively, the first and second moments of the random change
$\Delta {\mathbf{J}}_1$ suffered by the particle $1$ in a time $\Delta t$. The averaging is performed on the 
values of the angle variables of particles $2$ and on their action distribution functions. Equation 
(\ref{LAequation}) is recovered that way. In the calculation of the braking coefficient, 
small departures from uniform angular motion should be accounted for, as shown by \citet{Ecker} 
in a similar context.  The Fokker-Planck form of equation (\ref{LAequation}), although equivalent to it,
looks more complex than equation (\ref{LAequation}) itself
because the braking coefficient involves the derivative of a Dirac distribution.

\subsection{Accounting for strong collisions}
\label{strongcoll}

Equation (\ref{LAequation}) and its quasi-homogeneous limit, equation (\ref{LAequationhomograv}),
both result from a weak collision theory. Strong collisions involving  substancial
deviation of at least one of the colliding particles
are not adequately described.
This inappropriate description of the rare strong collisions can be fixed
by limiting the range of impact parameters to values larger than some critical limit $b_{cr}^{ab}$
which depends on the masses of the colliding species. This critical  impact parameter for
particles of species $a$ and $b$ is such that the typical kinetic energy
in their relative motion be equal to their interaction energy, that is:
\begin{equation}
\frac{G M}{R} \, \frac{m_a m_b}{m_a + m_b} = \frac{G m_a m_b}{b_{cr}^{ab}} \ \cdot
\end{equation}
Here $M$ is the total system's mass and $R$ a typical global size of it.
Were this cut to be omitted, the expressions of the coefficients in equations (\ref{LAequation}) and (\ref{LAequationhomograv})
would diverge logarithmically at large wavenumbers, where the response function $\varepsilon$
approaches unity.
This divergence results from the neglect of large deviations
in strong collisions.  A physically sound result is obtained
by limiting the ${\mathbf{K}}$ integration in equation (\ref{LAequationhomograv})
to the domain $\mid \! {\mathbf{K}}\! \mid < K_{cr}^{ab}$ where:
\begin{equation}
K_{cr}^{ab}  = \frac{2\pi}{b_{cr}^{ab}} = \frac{2\pi}{R} \ \frac{M}{m_a + m_b} \ \cdot
\end{equation}
Similarly the summations on the angle Fourier variables
${\mathbf{k}}_i$ ($i=$ 1 or 2) in equation (\ref{LAequation})
should be limited, in the term associated to species $a$ and $b$, to values such that the physical wavenumbers
along the quasi-intersecting orbits be smaller than $K_{cr}^{ab}$. This modulus of the physical wavenumber can be
crudely related to the dimensionless angle wavenumber by $K = k/R$, where $R$ is a typical global size of the system
and $k$ the modulus of the angle Fourier variable. Thus, the summation on  ${\mathbf{k}}_1$ and ${\mathbf{k}}_2$
in equation (\ref{LAequation}) should be limited to wave vectors, the modulus of which is bounded by:
\begin{equation}
\mid \! {\mathbf{k}}_i \! \mid  < \frac{ 2\pi \, M}{m_a + m_b} \ \cdot
\end{equation}

\bigskip

When solving equation (\ref{LAequation}),  the
secular evolution  of the response matrix $\varepsilon$, the 
system's collective potential $U({\mathbf{r}},t)$ and the
Fourier transform coefficients $\psi^{\alpha}_{\mathbf{k}}({\mathbf{J}})$ should be followed in time
together with the 1-body distributions. We return  
to this, in the case of spherical potentials, in section \ref{secsystemcouple}. 
Prior to that, let us discuss various limits and approximate forms
of equation (\ref{LAequation}) and show that,
as it should, it satisfies an H-theorem. 
The irreversibility stemms from the fact that information
is lost when the real issues of collisions are replaced in the equation by average ones, in particular by
averaging over the angles of the colliding particles.

\section{Limiting cases}
 
\subsection{Homogeneous limit}
\label{seclimitehomogene}

Although the limit of an homogeneous medium cannot be rigourously taken for a self gravitating system,
it is nevertheless possible to assume local homogeneity
at the price of artificially limiting the interaction distance between 
particles by cutting it at some characteristic size of the system. 
So doing, the effects of the collective dressing of the particles are still retained, albeit
less precisely, but the effects of the structure of the system are only sketchily
accounted for.

\bigskip

In this limit, the system is regarded as homogeneously filling
a large cubix box of side $L$, on the surface of which periodic
boundary conditions apply. This is the geometry considered by \citet{Weinberg93}.
Due to the assumed homogeneity, the collective force 
$\mathbf{F}^0$ vanishes and the unperturbed motion is rectilinear and uniform, 
whatever the state of relaxation of the system.
The action variables are then proportional to the components of the momentum $\mathbf{p}$ and the
angle variables are proportional to the components of the position $\mathbf{r}$. Since the
angles must be variables of period $2 \pi$, 
the angle vector must be ${\mathbf{w}} = 2 \pi {\mathbf{r}} /L$, which implies that
the action vector is ${\mathbf{J}} = L {\mathbf{p}}/2\pi$. The angle Fourier vector is
${\mathbf{k}} = L {\mathbf{K}} /2\pi$ where ${\mathbf{K}}$ is
the usual wave vector of  Fourier transforms with respect to position. 
The frequency vector ${\mathbf{\Omega}}$ is $2\pi {\mathbf{v}}/L$, so
that ${\mathbf{k}} \cdot {\mathbf{\Omega}} = {\mathbf{K}}   \cdot  {\mathbf{v}}$. 
The density-potential basis consists of functions proportional 
to complex exponentials, like $\exp(i\,{\mathbf{K}}\! \cdot\! {\mathbf{r}})$.  
A given element of the basis, $\alpha$ say,
is characterized by its wave vector ${\mathbf{K}}$. This can be accounted for
in the notation  by writing this wave vector as
${\mathbf{K}}_\alpha$, the corresponding angle wave vector being noted ${\mathbf{k}}_\alpha$.
The  density function and the potential of the element $\alpha$ of the basis are both
proportional to $\exp{(i\, {\mathbf{K}}_\alpha\! \cdot\! {\mathbf{r}}) }$. 
Their normalization factor must be such that
the biorthogonality relation (\ref{biortho}) be satisfied, the density
$D^\alpha({\mathbf{r}})$ and the potential $\psi^\alpha( {\mathbf{r}})$  being related by
equation (\ref{DeltaetPsi}). These constraints result in:
\begin{equation}
D^\alpha({\mathbf{r}}) =   \frac{\mid \! {\mathbf{K}}_\alpha  L\!\mid }{ 2\sqrt{\pi}\, L^{5/2}}
\ e^{i\, {\mathbf{K}}_\alpha \cdot {\mathbf{r}} }
\qquad \qquad \qquad 
\psi^\alpha({\mathbf{r}}) = - \, \frac{2\sqrt{\pi} }{L^{1/2} } \
\frac{e^{i\, {\mathbf{K}}_\alpha \cdot {\mathbf{r}} } }{ \mid\! {\mathbf{K}}_\alpha L \!\mid } \ \cdot
\end{equation}
The $\psi^\alpha_{ \mathbf{k} }$'s are the Fourier transforms of 
$\psi^\alpha({\mathbf{r}})$ with respect to the angles $\mathbf{w}$, namely:
\begin{equation}
\psi^\alpha_{\mathbf{k}} = - \frac{2 \sqrt{\pi}}{L^{1/2} } \ 
\frac{\delta ({\mathbf{k}}_\alpha - {\mathbf{k}})  }{ \mid\! {\mathbf{K}}_\alpha L \!\mid } \ , 
\label{psialphakhomograv}
\end{equation}
where $\delta({\mathbf{k}}_\alpha - {\mathbf{k}})$ is a triple Kronecker symbol. 
In this case the $\psi^\alpha_{\mathbf{k}}$'s do not depend on the actions
and remain fixed while the relaxation proceeds.
The response matrix $\varepsilon$,  calculated from its definition
(\ref{epsilonalphabeta}), is diagonal, its element $\alpha\alpha$ being given, for $\omega$ in the
upper half complex plane, by:
\begin{equation}
\varepsilon^{\alpha\alpha}(\omega) = 
1 - \sum_q \ \frac{4\pi \,G \, m_q^2}{\mid \! K_{\alpha}\! \mid^2} \ \, \int\!\! d^3\!p \ \,
\frac{ {\mathbf{K}}_{\alpha}\! \cdot\! {\mathbf{\nabla}}_{ {\mathbf{p}} } f^q({\mathbf{p}})  }{
\omega - {\mathbf{K}}_{\alpha} \! \cdot \! {\mathbf{v}} } \ \cdot
\label{epsilonhomograv}
\end{equation}
For real $\omega$, a $+i0$ should be added to the singular denominator.
Since $\alpha$ enters this relation by its wave vector $\mathbf{K}_\alpha$, $\varepsilon^{\alpha\alpha}(\omega)$
can be regarded as a function $\varepsilon({\mathbf{K}}_\alpha, \omega)$, or, more generally, 
as a function of a wave vector $\mathbf{K}$ and of a frequency $\omega$.
Because of the diagonality 
of the response matrix $\varepsilon$ and the simplicity of equation (\ref{psialphakhomograv}),
the writing of the kinetic equation (\ref{LAequation}) simplifies to:
\begin{equation}
\partial_t f^{a}({\mathbf{p}}) =
\sum_b \!  \int\!\! d^3\! p'
\ \  {\mathbf{\nabla}}_{ {\mathbf{p}}} \cdot\, \Big( \, 
{\overline{\overline{Q}}}_{ab} ({\mathbf{p}}, {\mathbf{p}}')
\, \cdot \, ({\mathbf{\nabla}}_{ {\mathbf{p}}} \! - {\mathbf{\nabla}}_{ {\mathbf{p}}'}) f^a({\mathbf{p}})
f^b({\mathbf{p}}') \, \Big) \ ,
\label{LAequationhomograv}
\end{equation}
where the tensor ${\overline{\overline{Q}}}_{ab}$ is defined by:
\begin{equation}
{\overline{\overline{Q}}}_{ab}({\mathbf{p}}, {\mathbf{p}}') = 2 G^2 m_a^2 m_b^2 \int\!\! d^3\!K \ \
\frac{ {\overline{\overline{ {\mathbf{K}} {\mathbf{K}} }}}  }{K^4} \ \
\frac{ \delta ({\mathbf{K}}\!\cdot \! ({\mathbf{v}} - {\mathbf{v}}' )) }{
\mid\! \varepsilon({\mathbf{K}}, {\mathbf{K}}\!\cdot\! {\mathbf{v}})\!\mid^2  } \ \cdot
\label{colltensorhomograv}
\end{equation}
Equations (\ref{LAequationhomograv})--(\ref{colltensorhomograv}) are identical to equation (29) 
of \citet{Weinberg93}  when the quasi homogeneity of the system
and associated absence of collective and external forces are accounted for.
Equation (\ref{LAequationhomograv}) can be written explicitly as a Fokker-Planck equation
in the form:
\begin{equation}
\partial_t f^{a}({\mathbf{p}}) =
- \, {\overline{{\mathbf{\nabla}}}}_{ {\mathbf{p}}}\cdot \Big({\overline{ \mathbf{A}}}_{a} f^a({\mathbf{p}})\Big)
+ {1 \over 2}\  {\overline{\overline{ {\mathbf{\nabla}}_{ {\mathbf{p}}} {\mathbf{\nabla}}_{ {\mathbf{p}}} }}} \, \colon \,
\Big({\overline{\overline{ {\mathbf{B}} }}}_{a} f^a({\mathbf{p}})\Big) \ , 
\label{formeFP}
\end{equation}
where the momentum drag and diffusion coefficients are:
\begin{eqnarray}
&&{\overline{{\mathbf{A}}}}_{a}({\mathbf{p}}) =  \sum_b 
\int \! \! d^3\!p'  f^b({\mathbf{p}}') \left(({\mathbf{\nabla}}_{ {\mathbf{p}}} - {\mathbf{\nabla}}_{ {\mathbf{p}}'})
\cdot {\overline{\overline{Q}}}_{ab}({\mathbf{p}}, {\mathbf{p}}')\right) \ , 
\label{Ahomogene}
\\
&& {\overline{\overline{ {\mathbf{B}} }}}_{a}({\mathbf{p}}) = 2 \sum_b 
\int \! \! d^3\!p' f^b({\mathbf{p}}')\ {\overline{\overline{Q}}}_{ab}({\mathbf{p}}, {\mathbf{p}}') \ \cdot
\label{Bhomogene}
\end{eqnarray}
For electrical instead of gravitational interactions, the  gravitational constant $G$ should be replaced 
by $1/4\pi \epsilon_o$ in MKSA units,
$\epsilon_o$ being the dielectric permittivity of vacuum. The electric force
between like charges being repulsive instead of attractive, the minus sign before the second term of equation 
(\ref{epsilonhomograv}) should be changed to a positive sign and
the masses replaced by the charges of the particles. 
Equation (\ref{LAequationhomograv}) then reduces to the Balescu-Lenard equation for
homogeneous and multispecies plasmas \citep{Babuel}. It
implicitly accounts for the screening effect, 
which is embodied in the dielectric function. The  
integral on wave vector space in equation (\ref{colltensorhomograv})
then need not be cut at small wave vectors because 
$\mid\! \varepsilon({\mathbf{K}}, {\mathbf{K}}\!\cdot {\mathbf{v}})\!\mid$ diverges as 
$\mid \!{\mathbf{K}}\!\mid$ approaches zero. 
For self gravitational systems the small $\mid \!{\mathbf{K}}\!\mid$
limit is unphysical, due to the absence of screening.
The distance between interacting particles 
is limited in this case by the inhomogeneity of the system, a
feature which is lost in the local approximation.
If one were to insist on the quasi-homogeneous approximation, the integration over wave vectors in
equation (\ref{colltensorhomograv}) would have to be artificially
limited from below to some minimum
modulus $K_{min} \sim 2\pi/R$, where $R$ is a characteristic size of the system. 
Little would then be gained over a more traditional Fokker-Planck approximation,
but for the fact that equation (\ref{LAequationhomograv}) still
accounts for the collective dressing of the colliding particles.

\subsection{Non-collective homogeneous limit}
\label{subseclimclassiq}

When these collective effects are themselves neglected, which amounts to take $\varepsilon = 1$  in equation
(\ref{colltensorhomograv}), 
the usual local Fokker-Planck equation (\ref{formeFP}) is recovered,
with braking and diffusion coefficients given by expressions (\ref{Ahomogene}) and (\ref{Bhomogene}), 
$\varepsilon$ now supposedly being equal to unity. As above,
the integral on wavevectors in equation (\ref{colltensorhomograv}) should limited to a lower cutoff
at $\mid \! {\mathbf{K}}\! \mid = K_{min}$, to account for the finite size of the system,
and to an upper cutoff $\mid \! {\mathbf{K}}\! \mid = K_{max}$, to
account for strong collisions (section \ref{strongcoll}). The coulomb logarithm is 
$\ln \Lambda$, where $\Lambda = K_{max}/K_{min}$.  
When $\varepsilon$ equals unity, the integration over wave vectors  in
equation (\ref{colltensorhomograv}) can easily be performed. The result, which involves the relative
velocity of the colliding particles  ${\mathbf{g}} = {\mathbf{v}} - {\mathbf{v}}'$, is: 
\begin{eqnarray}
&&{\overline{{\mathbf{A}}}}_{a}({\mathbf{p}}) = - 4\pi G^2 \,  \ln(\Lambda) \sum_b m_a m_b (m_a +m_b)
\int \! \! d^3\!p'  f^b({\mathbf{p}}') \ \, \frac{\mathbf{g}}{g^3} \ ,
\label{Asanseffetscoll}
\\
&& {\overline{\overline{ {\mathbf{B}} }}}_{a}({\mathbf{p}}) = + 4 \pi G^2 \,  \ln(\Lambda) \sum_b m_a^2 m_b^2 \,
\int \! \! d^3\! p' f^b({\mathbf{p}}') \, \ 
\frac{ {\overline{\overline{ {\mathbf{I}}  }}} \, g^2 - {\overline{\overline{ {\mathbf{g}}{\mathbf{g}} }}} }{g^3} \ ,
\label{Bsanseffetscoll}
\end{eqnarray}
where ${\overline{\overline{ {\mathbf{I}}  }}}$ is the unit second rank tensor.
This is identical to the Fokker-Planck equation presented, for example, in \citet{BinneyTremaine},
equations (8A.10). The collective dressing becomes important when $\mid\varepsilon\mid^{-2}$ in
equation (\ref{colltensorhomograv}) largely differs from unity.
As shown by \citet{Weinberg93},
this happens when the system is  not far from being unstable,
for example when its size becomes of order of the Jeans length,  
the complex zeroes of $\varepsilon({\mathbf{K}}, \omega)$ lying close, but below, the real axis.

\section{An H theorem}
\label{sectheoremH}

Equation (\ref{LAequation}) satisfies an H theorem  which states that the statistical entropy:
\begin{equation}
s(t) = -  \sum_a \! \int\! \! d^3w_1 \!\! \int\! \! d^3\!J_1 \,  f^a({\mathbf{w}}_1, {\mathbf{J}}_1, t) 
\ln\left(f^a({\mathbf{w}}_1, {\mathbf{J}}_1, t)\right) \ ,
\label{entropystat}
\end{equation}
increases with time.  Because the relaxing distribution functions
depend on actions only, the integral over angles reduces to a mere multiplication by a factor $8\pi^3$, so that:
\begin{equation}
\frac{d s(t)}{dt} \!= -  8\pi^3 \sum_a  \int\! \! d^3\!J_1 \ \Big(1 + \ln\left(f^a({\mathbf{J}}_1, t)\right)\Big) 
\ \Big(\partial_t f^a({\mathbf{J}}_1,t)\Big) \ \cdot
\label{evolentropystat}
\end{equation}
The time derivative of $f^a$ is given by equation (\ref{LAequation}) which
can be symmetrized by substituting to the first operator ${\mathbf{k}}_1\! \cdot\! {\mathbf{\nabla}}_{ {\mathbf{J}}_1}$
the operator 
${\mathbf{k}}_1\! \cdot\! {\mathbf{\nabla}}_{ {\mathbf{J}}_1} 
-\,  {\mathbf{k}}_2\! \cdot\! {\mathbf{\nabla}}_{ {\mathbf{J}}_2 }$. The 
contribution associated with the added operator $\! {\mathbf{k}}_2\! \cdot\! {\mathbf{\nabla}}_{ {\mathbf{J}}_2 }$
vanishes on integration over ${\mathbf{J}}_2$. This can be seen by using the flux divergence theorem
in action space, recognizing that the surface integral
over the boundary of the physical ${\mathbf{J}}_2$ domain vanishes. 
Indeed, the expression on the right of the first operator 
${\mathbf{k}}_1\! \cdot\! {\mathbf{\nabla}}_{ {\mathbf{J}}_1}$
in equation (\ref{LAequation}) represents, up to its sign, the flux in action space 
at ${\mathbf{J}}_1$ caused by collisions with particles having action 
${\mathbf{J}}_2$ or  the flux at ${\mathbf{J}}_2$  caused by collisions with particles of action ${\mathbf{J}}_1$.
The physical domain is limited in action space by a boundary at a finite distance and extends
to infinity in certain directions. The flux through the  boundary at finite distance vanishes because 
the action vector of no particle can 
evolve through this boundary from the physical to the unphysical domain.
The flux at infinity vanishes because $f^b({\mathbf{J}}_2)$ 
decreases fast enough.
This justifies the above-suggested substitution. 
The expression of $\partial_t f^a(1)$ given by equation (\ref{LAequation}),
modified as described, when inserted in equation (\ref{evolentropystat}), gives 
the following expression for $ds/dt$:
\begin{eqnarray}
&&\frac{d s(t)}{dt} =  - 64 \pi^7 \sum_a \sum_b \sum_{{\mathbf{k}}_1} \sum_{{\mathbf{k}}_2}  \int\! \! d^3\!J_1\!\! 
\int\! \! d^3\!J_2\  G^2 m_a^2 m_b^2 \Big(1 + \ln(f^a({\mathbf{J}}_1))\Big) \, \cdots
\nonumber \\
&&   \qquad \qquad \cdots \, \left({\mathbf{k}}_1\! \cdot\! {\mathbf{\nabla}}_{{\mathbf{J}}_1} \!
- \! {\mathbf{k}}_2\! \cdot\! {\mathbf{\nabla}}_{{\mathbf{J}}_2}\right) 
\delta({\mathbf{k}}_1\! \cdot\! {\mathbf{\Omega}}_1 - {\mathbf{k}}_2\! \cdot\! {\mathbf{\Omega}}_2)
 \left\vert{\cal{D}}_{{\mathbf{k}}_1 {\mathbf{k}}_2}({\mathbf{J}}_1, {\mathbf{J}}_2, {\mathbf{k}}_1\! \cdot\! {\mathbf{\Omega}}_1)\right\vert^{-2} 
\left({\mathbf{k}}_1\! \cdot\! {\mathbf{\nabla}}_{{\mathbf{J}}_1} \!
- \! {\mathbf{k}}_2\! \cdot\! {\mathbf{\nabla}}_{ {\mathbf{J}}_2 }\right) f^a({\mathbf{J}}_1)
f^b({\mathbf{J}}_2) \ \cdot
\end{eqnarray}
This expression is further symmetrized by combining it with the equivalent
expression obtained by exchanging species indices $a$ and $b$, 
actions ${\mathbf{J}}_1$ and ${\mathbf{J}}_2$ and Fourier variables ${\mathbf{k}}_1$ and ${\mathbf{k}}_2$.
The resulting expression is then integrated by parts, 
using the flux divergence theorem
in either ${\mathbf{J}}_1$ or ${\mathbf{J}}_2$ space.
As explained above, the contribution of the integral on the boundary of the action domain or at infinity vanishes. 
We are eventually left with the positive expression:
\begin{equation}
\frac{d s(t)}{dt} =  + 32 \pi^7 \sum_a \sum_b \sum_{{\mathbf{k}}_1} \sum_{{\mathbf{k}}_2}  \int\! \! d^3\!J_1\!\!
\int\! \! d^3\!J_2 \ \ \,  G^2 m_a^2 m_b^2 \ 
\frac{\delta({\mathbf{k}}_1\! \cdot\! {\mathbf{\Omega}}_1 - {\mathbf{k}}_2\! \cdot\! {\mathbf{\Omega}}_2) \ \ \left(
\left({\mathbf{k}}_1\! \cdot\! {\mathbf{\nabla}}_{ {\mathbf{J}}_1} \!
- \! {\mathbf{k}}_2\! \cdot\! {\mathbf{\nabla}}_{ {\mathbf{J}}_2 }\right) f^a({\mathbf{J}}_1) f^b({\mathbf{J}}_2) \right)^2  
}{
\left\vert{\cal{D}}_{{\mathbf{k}}_1 {\mathbf{k}}_2}({\mathbf{J}}_1, {\mathbf{J}}_2, {\mathbf{k}}_1\! \cdot\! {\mathbf{\Omega}}_1 )\right\vert^{2} \ 
f^a({\mathbf{J}}_1) f^b({\mathbf{J}}_2)} \ \cdot
\end{equation}
This establishes that the statistical entropy of 
the system is monotonically increasing. Since the entropy of a self gravitating system
of a given total mass and energy is not bounded from above \citep{BinneyTremaine},  
the increase of the statistical entropy does not
lead, as in homogeneous gases or plasmas, 
to a state of thermodynamic equilibrium. When the system becomes sufficiently centrally condensed,
a gravothermal instability develops \citep{Henon61, Antonov, LyndenBellWood}. 
The quenching of this instability by the formation of binaries 
is not described by equation (\ref{LAequation}), because the formation of  binary systems
results from triple collisions (that is, from third order correlations)
and is of the strong interaction type. 

\section{Evolution of a
spherical potential and basis Fourier coefficients} 
\label{secsystemcouple}

When the relaxation proceeds, the distribution functions $f^a({\mathbf{J}}, t)$ evolve according to equations 
(\ref{LAequation}). This causes a slow secular change in the 
average potential $U(\mathbf{r},t)$ and in the response matrix
$\varepsilon(\omega)$ (equation (\ref{epsilonalphabeta})). The basis potential functions $\psi^\alpha({\mathbf{r}})$ 
are time-independent, but their
Fourier transforms with respect to angles $\mathbf{w}$ are not
because they depend on the orbits of the particles, which
slowly change with the potential.  
The Fourier coefficients $\psi^\alpha_{{\mathbf{k}}}({\mathbf{J}})$ 
are then indirectly related to the slowly evolving potential $U({\mathbf{r}},t)$.

\bigskip

Thus, equation (\ref{LAequation}) is not an autonomous equation for
the distribution functions $f^a({\mathbf{J}}, t)$. 
The response matrix
$\varepsilon(\omega)$ of equation (\ref{epsilonalphabeta})
is a functional of those, which also depends 
on the angle Fourier transforms $\psi^{\alpha}_{{\mathbf{k}}}({\mathbf{J}}, t)$
as does the quantity $\cal{D}$  present in
equations (\ref{definitionD}) and (\ref{LAequation}).
The kinetic equation (\ref{LAequation}) must then be completed with equations describing 
the evolution in time of the average potential $U({\mathbf{r}},t)$
and of the angle Fourier coefficients $\psi^\alpha_{\mathbf{k}}({\mathbf{J}}, t)$ of the basis potentials. 
This aspect of the system's evolution is not considered by \citet{Chavanis}.
In this section, the time $t$ will be 
restaured, though only where necessary, in the list of arguments of functions.
The potential $U({\mathbf{r}},t)$ derives from the mass density:
\begin{equation}
D({\mathbf{r}},t) = \sum_a m_a \int\! d^3\!p \  f^a({\mathbf{r}}, {\mathbf{p}},t) \ \cdot
\label{densitemassefa}
\end{equation}
The corresponding gravitational potential is obtained from its expansion on the biorthogonal 
density-potential basis by equation (\ref{phiexpansiongene}). From appendix \ref{sphericalities}, it is found 
that its partial time-derivative is:
\begin{equation}
\partial_t U({\mathbf{r}},t) \!= - \!\sum_a 8 \pi^3 G m_a \, \psi^{\alpha}({\mathbf{r}}) \!\!  \int\!\! d^3\! J
\ \ \partial_t \Big(f^a({\mathbf{J}},t) \ (\psi^{\alpha}_{\mathbf{0}}({\mathbf{J}}, t))^* \Big) \ \cdot
\label{variationpot}
\end{equation}
Equation (\ref{variationpot}), as well as equations
(\ref{epsilonalphabeta}), (\ref{definitionD}) and
(\ref{LAequation}), call for an equation for the time-evolution of
the coefficients $\psi^{\alpha}_{{\mathbf{k}} }({\mathbf{J}}, t)$:
\begin{equation}
\psi^{\alpha}_{{\mathbf{k}} }({\mathbf{J}}, t) =  \int\! d^3\!\! w \  e^{- i {\mathbf{k}} \cdot {\mathbf{w}} }  \,
\psi^{\alpha}({\mathbf{r}}({\mathbf{w}}, {\mathbf{J}})) \ \cdot
\label{LATFdepsi}
\end{equation}
An explicit expression for these coefficients when the potential is spherical is derived in appendix \ref{sphericalities}, 
which also gives a summary of angle and action variables for a particle moving in a spherical potential.
In this case, the coefficients $\psi^{\alpha}_{{\mathbf{k}} }({\mathbf{J}}, t)$  vanish
when the wave vector ${\mathbf{k}}$ has non-vanishing $k_2$ or $k_3$ components.
The non-vanishing coefficients depend only on the radial $k_1$ component, hereafter noted $k$. 
For conciseness, the variables $\mathbf{J}$, which are mere parameters,
are omitted wherever possible.
After some calculations, we find that:
\begin{equation}
\psi^{\alpha}_{k}(t) = 8 \pi^2 \Omega_1(t)
\int_{r_P(t)}^{r_A(t)} \frac{\cos W_{k}(r,t) \ \, \psi^\alpha(r)\, dr}{\sqrt{2 (E(t) - U(r,t)) - J_2^2/r^2}} \ ,
\qquad {\mathrm{where}} \quad
W_{k}(r,t)  =
\int_{r_P(t)}^{r}\!
\frac{ k \, \Omega_1(t) \ \, dr'}{\sqrt{2(E(t) - U(r',t)) - {J_2^2}/{r^{'2}}}} \ \cdot
\label{W+k1k2}
\end{equation}
When the potential changes, the radial distances $r_P$ and $r_A$  of the periapse
and apoapse change accordingly: the bounds of the integrals in equations (\ref{W+k1k2}),
(\ref{Omegaspheriq}) and (\ref{Jspheriq}) are time-dependent. 
These integrals are singular, though convergent, the periapse and apoapse distances being
the zeroes of the square root denominator:
\begin{equation} 
q(r,t) = 2 (E(t) - U(r,t)) - {J_2^2}/{r^2} \ \cdot
\label{p}
\end{equation}
These zeroes are simple when the orbit is not circular and merge into a double zero
when it is. This latter situation can be dealt with by a limit process, in which
simple zeroes $r_P$ and $r_A$ are made to converge to eachother. We then assume 
that $r_P$ and $r_A$ are simple zeroes. An index $P$ or $A$ denotes
the value of a function of $r$ at $r_P$ or $r_A$ respectively.
Integrals like: 
\begin{equation}
I(t) = \int_{r_P(t)}^{r_A(t)} \! dr \ \, \frac{m(r,t)}{\sqrt{q(r,t)}} \ ,
\label{singinteg}
\end{equation}
or similar ones can be expressed in terms of a variable $\xi$, the values of which 
remain constant at the changing periapse and apoapse. This variable is defined by:
\begin{equation}
r = r_P(t) + \xi \, (r_A(t) - r_P(t)) \ \cdot
\label{changerxi}
\end{equation}
To each of these two types of radial variables, $r$ or $\xi$, 
a time variable, $t$ or $\tau$, can be associated, it being meant that $t \equiv \tau$.
This introduces two types of time derivatives: $\partial_t$, which is at constant $r$,
and $\partial_\tau$, which is at constant $\xi$. Ordinary time derivatives with respect to $\tau$ and $t$ are identical
and are denoted by a dot. Partial time derivatives with respect to $\tau$ and $t$
differ and are related by:
\begin{equation}
\partial_\tau = \partial_t + 
\left( {\dot{r}}_A \left(\frac{r - r_P}{r_A - r_P}\right) + {\dot{r}}_P \left(\frac{r_A -r}{r_A - r_P}\right)\right) \, \partial_r \ \cdot
\label{partialtaut}
\end{equation}
The partial derivatives with respect to $r$ and $\xi$ are simply proportional:
$ \partial_\xi =  (r_A - r_P)\, \partial_r$.
At the periapse or apoapse the function $q(r,t)$
vanishes, whatever $\tau$. Hence, $\partial_\tau q = 0$ at these points.
Differentiating the equation $q(r,t) = 0$,  ${\dot{r}}_P$ and ${\dot{r}}_A$ are found: 
\begin{equation}
{\dot{r}}_P = - \, \frac{\partial_t q(r_P,t) }{\partial_r q(r_P,t)}  \,
\qquad  \qquad \qquad \qquad \qquad  
{\dot{r}}_A = - \, \frac{\partial_t q_(r_A,t) }{\partial_r q (r_A,t)} \ \cdot
\label{dotrPA}
\end{equation}
The partial derivatives of $q(r,t)$ (equation (\ref{p})) are:
\begin{equation}
\partial_t q(r,t) = 2( {\dot{E}} - \partial_t U(r,t)) \ ,
\qquad \qquad \qquad  
\partial_r q(r,t) = 2 \left(\frac{J_2^2}{r^3} - \partial_r U(r,t) \right) \ \cdot
\label{derq}
\end{equation}
It can be checked from equations (\ref{dotrPA}) and (\ref{partialtaut}) 
that $(\partial_\tau q)(r_P,t) = \partial_\tau q (r_A,t) = 0$. 
Equation (\ref{partialtaut}) implies that:
\begin{equation}
\partial_\tau r = {\dot{r}}_P \left(\frac{r_A -r}{r_A - r_P}\right) + {\dot{r}}_A \left(\frac{r - r_P}{r_A - r_P}\right)
\ , \qquad \qquad \qquad
\partial_\tau q = \partial_t q + 
\left( {\dot{r}}_P \left(\frac{r_A -r}{r_A - r_P}\right) + {\dot{r}}_A \left(\frac{r - r_P}{r_A - r_P}\right)\right)
\, \partial_r q  \ \cdot
\label{partialtauq}
\end{equation}
The time-derivative of $I(t)$ (equation (\ref{singinteg})) is found by changing the variable $r$ to $\xi$:
\begin{equation}
{\dot{I}} = \left(\frac{{\dot{r}}_A - {\dot{r}}_P}{r_A - r_P}\right) \, I 
+ \int_{r_P(t)}^{r_A(t)} \frac{m(r,t) \, dr}{\sqrt{q(r,t)}} 
\left(\frac{\partial_\tau m}{m(r,t)}- {1 \over 2} \, \frac{\partial_\tau q}{q(r,t)} \right) \ \cdot
\label{deltaIderivtau}
\end{equation}
It is important to note that the last term in the parenthesis of the integral in equation (\ref{deltaIderivtau}) is regular  since
the numerator, $\partial_\tau q$, vanishes at $r_P$ and $r_A$, where $q(r,t)$ does. The right hand side
of equation (\ref{deltaIderivtau}) then consists of convergent integrals.  
When this method is used to calculate the time derivatives of $E(t)$ and $\Omega_1(t)$ 
from equations (\ref{Jspheriq}) and (\ref{Omegaspheriq}), the following results are obtained:
\begin{equation}
{\dot{E}} = \frac{\Omega_1}{\pi} \  \int_{r_P(t)}^{r_A(t)} dr \ \frac{\partial_t U(r,t)}{\sqrt{q(r)}}\ ,
\qquad \qquad 
{\dot{\Omega}}_1 = - \Omega_1 \ \left(\frac{{\dot{r}}_A - {\dot{r}}_P}{r_A - r_P}\right)
+ \frac{\Omega_1^2}{2\pi}  \int_{r_P(t)}^{r_A(t)} \frac{dr}{\sqrt{q(r,t)}} \ \frac{\partial_\tau q}{q(r,t)} \ \cdot
\label{dotOmega1}
\end{equation}
The same method is used to calculate $\partial_\tau W_k$:
\begin{equation}
\partial_\tau W_k (r,t) = \frac{{\dot{\Omega}}_1}{\Omega_1} \ W_k (r,t) +
\left(\frac{{\dot{r}}_A - {\dot{r}}_P}{r_A - r_P}\right) W_k (r,t)
-  \frac{k_1 \Omega_1}{2} \! \int_{r_P(t)}^{r}\!  \frac{dr'}{\sqrt{q(r',t)}}
\ \, \frac{\partial_\tau q(r',t)}{q(r',t)} \ \cdot
\label{dottauWk}
\end{equation}
The angle Fourier coefficient $\psi^\alpha_{k}({\mathbf{J}}, t)$ is 
given by equation (\ref{W+k1k2}), which is of a form similar to equation (\ref{singinteg}).
Using the general result (\ref{deltaIderivtau}), the time derivative of $\psi^\alpha_{k}({\mathbf{J}}, t)$ is found to be:
\begin{equation}
 {\dot{\psi}}^\alpha_{k}({\mathbf{J}}, t) = 
\frac{{\dot{\Omega}}_1}{\Omega_1} \ \psi^\alpha_{k}
+ \left(\frac{{\dot{r}}_A - {\dot{r}}_P}{r_A - r_P}\right) \ \psi^\alpha_{k}
- 8\pi^2 \Omega_1\! \int_{r_P(t)}^{r_A(t)}\!  \frac{\psi^\alpha(r) \, dr}{\sqrt{q(r,t)}} 
 \left(\! \sin\! W_k (r,t) \, \partial_\tau W_k 
- \cos\! W_k (r,t) \left(\frac{\psi^{'\alpha}(r) \, \partial_\tau r}{\psi^\alpha(r)} 
- \frac{\partial_\tau q}{2 q(r,t)}\right)\! \right) \cdot
\label{dotdepsialphak}
\end{equation}
Equation (\ref{dotdepsialphak}) describes the time-evolution of the Fourier 
coefficients $\psi^\alpha_k({\mathbf{J}}, t)$.
The auxiliary $\tau$-derivatives which enter this
equation are given by equations (\ref{dottauWk}), (\ref{dotOmega1}), (\ref{partialtauq}),
(\ref{dotrPA}) and (\ref{derq}). The other quantities entering equation
(\ref{dotdepsialphak}) are expressed in terms of the potential by equations (\ref{W+k1k2}),
(\ref{p}), (\ref{periapoequation}) and  (\ref{Omegaspheriq}). 
All these relations eventually link
the time-evolution of $\psi^\alpha_{k}$ to the time-evolution of 
$U(r,t)$, which is itself described by equation (\ref{variationpot}). 

\bigskip

Equations (\ref{LAequation}), (\ref{variationpot}) and (\ref{dotdepsialphak}) 
form the system of coupled
equations for the distribution functions $f^a({\mathbf{J}},t)$, the average potential $U(r,t)$ and
the angle Fourier coefficients $\psi^\alpha_k({\mathbf{J}}, t)$ 
that we have been seeking for in this section. This system involves the auxiliary
equations mentioned above, as well as equations (\ref{epsilonalphabeta})--(\ref{definitionD}).

\section{Conclusions}
\label{secconclu}

Kinetic equations for the collisional evolution of the constituents of self-gravitating inhomogeneous systems 
have been derived. These equations (\ref{LAequation}) surpass in consistency 
the usual Fokker-Planck equations (\ref{formeFP})~--~(\ref{Asanseffetscoll})~--~(\ref{Bsanseffetscoll}). 
The latter are unsatisfactory
from a principle point of view, being local and non-collective. 

\bigskip

By contrast, the proposed equations fully account for the system's
inhomogeneity and for the collective gravitational dressing
of the colliding particles.

\bigskip 

Equations (\ref{LAequation}) describe the evolution of distribution functions
in action and angle space, which is possible when the hamiltonian associated with the 
average potential is integrable. 

\bigskip

Physically, these equations describe the evolution of the distribution functions in action space as a result 
of the weak gravitational noise caused by the discreteness of the particles,
dressed with the polarization clouds that their own gravity induces in the system. 
This gravitational polarization is
accounted for in equation (\ref{LAequation}) in a manner
that is fully consistent with the distribution functions, as they are at the moment. 

\subsection{Properties of the kinetic equation}
\label{subsecproperties}

Equation (\ref{LAequation}) is the sum of a second order derivative term with respect to actions
and of a first order one. It therefore basically is of the Fokker-Planck type, although it
is definitely simpler in the form of expression (\ref{LAequation}). The diffusion coefficient 
involved depends on the 1-body distributions themselves, in particular through the
factor $\mid \!{\cal{D}}\!\mid^{-2}$ which represents the effect of the dressing of the colliding
particles by the gravitational polarization induced around them by
their own influence.

\bigskip

Unlike in electrical plasmas, the polarization dressing
in self-gravitational systems does not cause any screening of the
interaction, which remains  effective even between distant particles. The mutual distance
of such particles is limited only by the finite size of the system.
Were the gravitational influence of particles on their surrounding to be neglected, the response matrix
$\varepsilon$ (equation (\ref{epsilonalphabeta})) would reduce to unity and the
coefficients of the corresponding Fokker-Planck 
kinetic equation would simply be averages by the distribution functions of functions
of velocity, as in equations (\ref{Asanseffetscoll})~--~(\ref{Bsanseffetscoll}).  

\bigskip

It is apparent from the developments of appendix \ref{grossesmagouilles},
which lead to equation (\ref{LAequation}), that the
${\mathbf{k}}$ component in angle Fourier space of
the gravitational polarization response given to a particle
has frequency  $\omega = {\mathbf{k}}\! \cdot \! {\mathbf{\Omega}}$. This means that the polarization
cloud which accompanies a particle forms a structure in angle space which vary as ${\mathbf{w}} - {\mathbf{\Omega}}t$: it
corotates in angle with that particle.

\bigskip

The presence of the Dirac function
$\delta({\mathbf{k}}_1\!\cdot  {\mathbf{\Omega}}_1 - {\mathbf{k}}_2\!\cdot  {\mathbf{\Omega}}_2)$ in equation
(\ref{LAequation}) indicates that particles interact resonantly. This certainly is an important
physical property of remote interactions, for which the components of the angle wave vectors ${\mathbf{k}}_1$ and
${\mathbf{k}}_2$ must be small. For closer encounters, the modulus of these wave vectors is larger
and the resonance condition ${\mathbf{k}}_1\cdot  {\mathbf{\Omega}}_1 = {\mathbf{k}}_2\cdot  {\mathbf{\Omega}}_2$
becomes less selective, being more easily satisfied.

\bigskip

The correlation function has been  calculated on the basis of a linearized theory, which
is justified by the weakness of the average interactions in this many-body system. This means
that the trajectories of the particles during the collision are regarded as being the unperturbed trajectories.
Similarly, the gravitational polarization cloud around any one of the colliding particles
is calculated as if the partner in the collision were not present: equation (\ref{LAequation}) is still
a weak collision approximation.
A cutoff at small impact parameters is therefore needed to account for
the rare strong collisions.

\bigskip

Equation (\ref{LAequation}) takes full account of the inhomogeneity  of the system, which is embodied in
the dependence of the distribution functions
on the actions $\mathbf{J}$'s. 
It requires no artificial cutoff
at large impact parameters.
The details of the trajectories followed by the particles in the
present gravitational potential are also fully accounted for, being implicit in  the relations which
link the angle and action variables to the position and momentum ones. These relations depend on the
actual global gravitational potential of the system, which slowly evolves in time together
with the distribution functions.

\bigskip

The density-potential basis functions
$\psi^{\alpha}({\mathbf{r}})$ are choosen at the beginning of the calculation
once and for all, but
their angle Fourier transforms $\psi^{\alpha}_{\mathbf{k}}({\mathbf{J}})$, which
depend on the actual trajectories of the particles, change with time
because the trajectory of a particle of  given actions slowly evolves with
the general potential of the system as the relaxation proceeds.
As long as it suffers no collision,
a given particle keeps its vector  ${\mathbf{J}}$ fixed because the actions are adiabatic invariants.
Collisions, however, cause a secular evolution of the functions $f^a({\mathbf{J}})$, which is exactly what
equation (\ref{LAequation}) describes.

\bigskip

\noindent
The description of particle motions 
is made simple by the use of action and angle variables.
Their complexity is embodied in the supposedly known relation between 
position and momentum variables and action and angle variables. The usefulness of 
equation (\ref{LAequation}) is therefore limited to systems for which this relation can be established, 
either analytically or, possibly, numerically \citep{PichonCannon, McMillanBinney}.

\bigskip

\noindent
While the relaxation proceeds, the gravitational potential and the orbits
of the particles evolve. As a result, the kinetic equation 
must be completed by evolution equations for the potential and for other relevant quantities. Section 
\ref{secsystemcouple} establishes, for spherical systems, this set of coupled equations.

\subsection*{Acknowledgements}
\label{acknowledgements}

I am indebted to C. Pichon and D. Aubert for awaking my interest for this problem.
Their work \citep{PichonAubert}
introduced me to the tools which make it possible
to extend the methods of plasma physics to these systems. I thank the referee, C. Pichon, 
for many suggestions which helped to improve the first version of this paper. I also thank 
the Observatoire Astronomique and the Universit\'e de Strasbourg for accepting me
as professor emeritus.

{}

\begin{appendix}

\section{Derivation of the kinetic equation}
\label{grossesmagouilles}

Equation (\ref{Colloperateurbrut}) must be expressed in terms of angle and action variables,
using the adopted density-potential basis. 
The integration on the dynamical state of the particle of species $b$ 
should be carried out first.
The integral over the variables $2$ in the second line 
of eq.(\ref{Colloperateurbrut}) is 
similar to equation (\ref{laformegenerale}), with $\lambda = G m_a m_b$, 
$M(2) = {\cal{G}}^{b}_q(2,2',\omega')$ and
$N(1) = {\cal{G}}^a_p(1, 1', \omega)$ and is thus 
expressed 
in angle and action variables  
by equation (\ref{expression14quasifinale}).
Since $M(2)$ is as in equation (\ref{VlasovlinpourG1Lapl}), the $a_\alpha$ coefficients are 
those of equation (\ref{adeFtransformG}), the species indices being now $b$ and $q$ instead of $c$ and $p$ 
and the parameters being $2', \omega'$
instead of  $1', \omega$. This leads to the following change in
equation (\ref{Colloperateurbrut}):
$$\displaylines{
\sum_b \!\int\! d2 \,
\left({\mathbf{F}}_{ab}(1,2) \cdot {\mathbf{\nabla}}_{{\mathbf{p}}_1}\right)\, {\cal{G}}^a_p(1,1', \omega) \,
{\cal{G}}^b_q(2,2',\omega') \ \ =
\hfill \cr} $$ 
\begin{equation}
i \, G m_a m_q \sum_{\mathbf{k}_2}\sum_{\mathbf{k}'_2} \, 
\left(\varepsilon^{-1}(\omega')\right)^{\alpha \beta}
\psi^{\beta*}_{\mathbf{k}'_2}({\mathbf{J}}'_2) 
\ \ \frac{e^{i({\mathbf{k}}_2 \cdot {\mathbf{w}}_1 - {\mathbf{k}}'_2 \cdot {\mathbf{w}}'_2)}
  }{
\omega' - {\mathbf{k}}'_2 \cdot  {\mathbf{\Omega}}'_2 } \ \
\left(\Big( \psi^\alpha_{\mathbf{k}_2}({\mathbf{J}}_1) \ i {\mathbf{k}_2}\cdot {\mathbf{\nabla}}_{{\mathbf{J}}_1} 
- ({\mathbf{\nabla}}_{{\mathbf{J}}_1} \psi^\alpha_{\mathbf{k}_2}({\mathbf{J}}_1)) \cdot 
{\mathbf{\nabla}}_{{\mathbf{w}}_1}\Big)\  {\cal{G}}^a_p (1, 1', \omega)\right) \ \cdot
\label{ligne2deO2}
\end{equation}
Note that, as a general rule, operators act 
on everything on their right, up to the end of the expression or to a closing delimiter.
Using the relation (\ref{ligne2deO2}), the collision operator (\ref{Colloperateurbrut}) can be written as:
\begin{eqnarray}
{\cal{C}}^a(f) &=& - \sum_{p,\, q} \int_0^\infty \!\!\! d\tau
\int\!\!\! d1'\! \int\!\!\! d2' \!
\int_B \! \frac{d\omega}{2\pi} \int_{B'} \! \frac{d\omega'}{2\pi}  e^{-i \, (\omega + \omega') \tau}
\sum_{\mathbf{k}_2}\sum_{\mathbf{k}'_2} \ 
i\ G m_a m_q \, (\varepsilon^{-1}(\omega'))^{\alpha \beta} 
\psi^{\beta*}_{ {\mathbf{k}}'_2}({\mathbf{J}}'_2) 
\ \, \frac{ e^{ i({\mathbf{k}}_2 \cdot {\mathbf{w}}_1 - {\mathbf{k}}'_2 \cdot {\mathbf{w}}'_2) }
  }{
\omega' - {\mathbf{k}}'_2 \cdot  {\mathbf{\Omega}}'_2 }
\nonumber \\
&& \Big(\, \Big( \psi^\alpha_{{\mathbf{k}}_2}({\mathbf{J}}_1) \ i {\mathbf{k}}_2\cdot {\mathbf{\nabla}}_{ {\mathbf{J}}_1}
- ({\mathbf{\nabla}}_{ {\mathbf{J}}_1 } \psi^\alpha_{\mathbf{k}_2}({\mathbf{J}}_1)) \cdot
{\mathbf{\nabla}}_{{\mathbf{w}}_1} \Big) {\cal{G}}^a_p (1, 1', \omega)\, \Big) \ \ 
\Big(\,
{\mathbf{F}}_{pq}(1', 2')  \cdot 
\left({\mathbf{\nabla}}_{{\mathbf{p}}'_2} - {\mathbf{\nabla}}_{{\mathbf{p}}'_1} \right) f^p(1') f^q(2') 
\, \Big) \ \cdot
\label{Colloperateurtransfo1}
\end{eqnarray}
The 1-body propagator ${\cal{G}}^a_p (1, 1', \omega)$ is then Fourier-expanded with respect to
both angles ${\mathbf{w}}_1$ and ${\mathbf{w}}'_1$ according to equation (\ref{defexpansFourierw}) and this expansion 
is inserted in equation (\ref{Colloperateurtransfo1}). It then appears that 
${\cal{C}}^a(f)$ depends on ${\mathbf{w}}_1$ 
as $\exp(i ({\mathbf{k}}_1 + {\mathbf{k}_2}) \cdot {\mathbf{w}}_1)$. Since 
during relaxation $f^a(1)$ remains a function of ${\mathbf{J}}_1$ only, 
it is possible to average over ${\mathbf{w}}_1$ without loss of
information, which brings a Kronecker factor 
$\delta({\mathbf{k}}_1 + {\mathbf{k}}_2)$, such that ${\mathbf{k}}_1 = - {\mathbf{k}}_2 = {\mathbf{k}}$.
The angle-averaged form of the collision operator is:
\begin{eqnarray}
{\cal{C}}^{\, a}\! (f) &=& -  \sum_{p,\, q} \sum_{ {\mathbf{k}}}
\int_0^\infty \! d\tau
\! \int\! d1'\! \int\! d2' \!
\int_B \! \frac{d\omega}{2\pi} \! \int_{B'} \! \frac{d\omega'}{2\pi} e^{-i \, (\omega + \omega') \tau} 
\, {\mathbf{k}}\! \cdot\! {\mathbf{\nabla}}_{ {\mathbf{J}}_1 }  \  \Bigg[\,  \psi^\alpha_{-{\mathbf{k}}}(1)
 (\varepsilon^{-1}(\omega'))^{\alpha \beta} 
\nonumber \\
&& \left(\sum_{ {\mathbf{k}}'_2 } \, G m_a m_q  
\frac{ \psi^{\beta*}_{\mathbf{k}'_2}(2') }{
\omega' - {\mathbf{k}}'_2 \cdot  {\mathbf{\Omega}}'_2 } 
e^{-i\,{\mathbf{k}}'_2 \cdot {\mathbf{w}}'_2}\right) \! 
\left(\sum_{ {\mathbf{k}}'_1 } e^{i\, {\mathbf{k}}'_1 \cdot {\mathbf{w}}'_1} \, 
G^{ap}_{ {\mathbf{k}} {\mathbf{k}}'_1}(1,\! 1'\! , \omega)
\ {\mathbf{F}}_{pq}(1',\! 2')\!  \cdot \!({\mathbf{\nabla}}_{{\mathbf{p}}'_2} \!-\!
{\mathbf{\nabla}}_{ {\mathbf{p}}'_1 })  f^p(1') f^q(2') \right) \, \Bigg] \ \cdot
\label{Colloperateurmoyen}
\end{eqnarray}
The following calculations are somewhat similar to those carried out for an homogeneous plasma by \citet{Ichimaru}.
The expression (\ref{Colloperateurmoyen}) can be split into two parts, one, 
${\cal{C}}^{\, a}_1 (f)$, being
associated with the operator ${\mathbf{\nabla}}_{{\mathbf{p}}'_1}$ in the last parenthesis
and the other, ${\cal{C}}^{\, a}_2 (f)$, being associated with
the operator ${\mathbf{\nabla}}_{{\mathbf{p}}'_2}$, so that:
\begin{equation}
{\cal{C}}^{\, a}\! (f) = {\cal{C}}^{\, a}_1\! (f) + {\cal{C}}^{\, a}_2\! (f) \ \cdot
\end{equation} 
The terms
${\mathbf{F}}_{pq}(1',\! 2') \cdot \! {\mathbf{\nabla}}_{ {\mathbf{p}}'_2} f^q(2')$ and
${\mathbf{F}}_{pq}(1',\! 2') \cdot \! {\mathbf{\nabla}}_{ {\mathbf{p}}'_1} f^p(1')$, multiplied by
other functions of $1'$ and $2'$ respectively, are subject to
an integration over these variables. The structure of these
expressions being similar to equation (\ref{laformegenerale}), 
they are expressed as in equation (\ref{PoissonbracketwJ}), noting 
that $f^p(1')$ and $f^q(2')$ do not depend on angles. 
The coefficients $a_\alpha$ of the development on the density-potential basis (equation (\ref{phiexpansiongene}))
which enter equation (\ref{PoissonbracketwJ})
are calculated from equation (\ref{coeffadephib}), with appropriate $M$ functions. 
Integration over angles ${\mathbf{w}}'_1$ or ${\mathbf{w}}'_2$ can then easily be carried out.
We are left with:
\begin{eqnarray}
&&{\cal{C}}^{\, a}_1\!(f) \, = \, - i  \sum_{p}\! \sum_{q} \int_0^\infty \!\!\! d\tau\!
\int_B \! \frac{d\omega}{2\pi} \! \int_{B'} \! \frac{d\omega'}{2\pi} e^{-i \, (\omega + \omega') \tau}
\sum_{ {\mathbf{k}}}  (8 \pi^3 G)^2 m_a m_p m_q^2
\, ({\mathbf{k}}\! \cdot\! {\mathbf{\nabla}}_{ {\mathbf{J}}_1 })  \ \Bigg[ \,   \psi^\alpha_{-{\mathbf{k}}}(1)
 (\varepsilon^{-1}(\omega'))^{\alpha \beta}
\nonumber \\
&& \left( \sum_{\mathbf{k}'_1}  \int\!\! d^3\!J'_1 \ \
G^{ap}_{ {\mathbf{k}} {\mathbf{k}}'_1}(1,\! 1'\!, \omega) \psi^\gamma_{ - {\mathbf{k}}'_1}(1')
\left({\mathbf{k}}'_1 \cdot {\mathbf{\nabla}}_{ {\mathbf{J}}'_1} f^p(1')\right) \right)
\left(  \sum_{{\mathbf{k}}'_2}\int\!\! d^3\!J'_2 \ \
\frac{ \psi^{\beta*}_{\mathbf{k}'_2}(2')  \psi^{\gamma*}_{-\mathbf{k}'_2}(2') \  f^q(2') }{
\omega' - {\mathbf{k}}'_2 \cdot  {\mathbf{\Omega}}'_2 } \right) \, \Bigg] \ ,
\label{Colloperateur1}
\end{eqnarray}
\newpage
\begin{eqnarray}
&&{\cal{C}}^{\, a}_2\!(f) \, = \, +i  \sum_{p}\! \sum_{q} \int_0^\infty \!\!\! d\tau\!
\int_B \! \frac{d\omega}{2\pi} \! \int_{B'} \! \frac{d\omega'}{2\pi} e^{-i \, (\omega + \omega') \tau}
\ \sum_{ {\mathbf{k}}}  (8 \pi^3 G)^2 m_a m_p m_q^2  
\, ({\mathbf{k}}\! \cdot\! {\mathbf{\nabla}}_{ {\mathbf{J}}_1 })  \ \Bigg[\,  \psi^\alpha_{-{\mathbf{k}}}(1) 
 (\varepsilon^{-1}(\omega'))^{\alpha \beta} 
\nonumber \\
&& \left( \sum_{ {\mathbf{k}}'_1} \int\!\! d^3\!J'_1 \ \  G^{ap}_{ {\mathbf{k}} {\mathbf{k}}'_1}(1,\! 1'\!, \omega)
\psi^{\gamma*}_{\mathbf{k}'_1}(1') f^p(1')\right)
\left( \sum_{{\mathbf{k}}'_2}\int\!\! d^3\!J'_2 \ \
\psi^{\beta*}_{\mathbf{k}'_2}(2')  \psi^\gamma_{\mathbf{k}'_2}(2')
\, \frac{{\mathbf{k}}'_2\! \cdot\! {\mathbf{\nabla}}_{ {\mathbf{J}}'_2} f^q(2')  }{
\omega' - {\mathbf{k}}'_2 \cdot  {\mathbf{\Omega}}'_2 } \right) \, \Bigg] \ \cdot
\label{Colloperateur2}
\end{eqnarray}
The integrations over $\tau$ and $\omega'$ which appear in equations 
(\ref{Colloperateur1})~--~(\ref{Colloperateur2})
are of the general form~
\begin{equation}
h(\omega) = \int_0^\infty \!\!\! d\tau\!  \int_{B'} \! \frac{d\omega'}{2\pi} e^{-i \, (\omega + \omega') \tau}
f(\omega)\, g(\omega') \ \cdot
\label{formintegraleomegaprim}
\end{equation}
The integration over $\tau$ is regular, and straightforward,  when $\omega + \omega'$ has a negative imaginary part.  Otherwise 
the result must be obtained by analytical continuation. This  means that, whatever $\omega$:
\begin{equation}
h(\omega) = \int_{B'} \! \frac{d\omega'}{2\pi}  \frac{-i}{\omega + \omega'} f(\omega)\, g(\omega') \ \cdot
\end{equation}
The contour $B'$ passes above all singularities of $g(\omega')$. For a stable system these are all 
below or on the real axis. When $\omega$ is low enough in the lower half complex plane $C^-$  for
$-\omega$ to be above $B'$, the integration on $\omega'$ can be carried out by closing the contour $B'$
at infinity in the upper half complex plane $C^+$, using the theorem of residues at the 
unique singularity in the closed up contour, which is at $\omega' = - \omega$. The closing of $B'$
in $C^+$ is possible because in the present case (see equations (\ref{Colloperateur1}~--~\ref{Colloperateur2}))
$g(\omega')/(\omega + \omega')$ decreases at infinity as $\mid \omega'\mid^{-2}$, which means that for such $\omega$'s, 
$h(\omega) = f(\omega)\, g(-\omega)$. Analytical continuation extends this result to other $\omega$'s.
The two parts of the collision operator then reduce to:
\begin{eqnarray}
&&{\cal{C}}^{\, a}_1\!(f) = \!-i \!  \int_B \! \frac{d\omega}{2\pi} \!\sum_{ {\mathbf{k}}} \
(8 \pi^3 G)^2 m_a
({\mathbf{k}}\! \cdot\! {\mathbf{\nabla}}_{ {\mathbf{J}}_1 })  \ \Bigg[ \,  \psi^\alpha_{-{\mathbf{k}}}(1) \,
(\varepsilon^{-1}(-\omega))^{\alpha \beta}
\nonumber \\
&& \left( \sum_p m_p \, \sum_{ {\mathbf{k}}'_1} \int\!\! d^3\!J'_1  \  \
G^{ap}_{ {\mathbf{k}} {\mathbf{k}}'_1}(1,\! 1'\!, \omega)
\, \psi^\gamma_{ - {\mathbf{k}}'_1}(1') \,
\left({\mathbf{k}}'_1\! \cdot\! {\mathbf{\nabla}}_{ {\mathbf{J}}'_1} f^p(1')\right) \right)
\left( \sum_q  m_q^2 \sum_{ {\mathbf{k}}'_2 }  \int\!\! d^3\!J'_2\ \
\frac{  \psi^{\beta*}_{{\mathbf{k}}'_2}(2') \psi^{\gamma*}_{ - {\mathbf{k}}'_2}(2') f^q(2')  }{
-\omega - {\mathbf{k}}'_2 \cdot  {\mathbf{\Omega}}'_2 }\right) \, \Bigg] \ ,
\label{Colloperateur1integromega'}
\end{eqnarray}
\begin{eqnarray}
&&{\cal{C}}^{\, a}_2\!(f)\! =\! + i \! 
\int_B \! \frac{d\omega}{2\pi} \! \sum_{ {\mathbf{k}}}
(8 \pi^3 G)^2 m_a  
({\mathbf{k}}\! \cdot\! {\mathbf{\nabla}}_{ {\mathbf{J}}_1 })  \ \Bigg[ \,  \psi^\alpha_{-{\mathbf{k}}}(1) \,
 (\varepsilon^{-1}(-\omega))^{\alpha \beta}  
\nonumber \\
&& \left(\sum_p \, m_p \sum_{ {\mathbf{k}}'_1} \int\!\! d^3\!J'_1 \ \ G^{ap}_{ {\mathbf{k}} {\mathbf{k}}'_1}(1,\! 1'\!, \omega) 
\psi^{\gamma*}_{\mathbf{k}'_1}(1')  f^p(1')\right)
\left(\sum_q m_q^2 \sum_{{\mathbf{k}}'_2 } \int\!\! d^3\!J'_2 \ \
\psi^\gamma_{\mathbf{k}'_2}(2') \psi^{\beta*}_{\mathbf{k}'_2}(2')
\ \frac{ \left({\mathbf{k}}'_2 \cdot \! {\mathbf{\nabla}}_{ {\mathbf{J}}'_2} f^q(2')\right)
}{
-\omega - {\mathbf{k}}'_2 \cdot  {\mathbf{\Omega}}'_2 }\right) \, \Bigg] \ \cdot
\label{Colloperateur2integromega'}
\end{eqnarray}
The last parenthesis in the second line of equation (\ref{Colloperateur2integromega'}) is 
$(\delta^{\beta \gamma} - \varepsilon^{\beta \gamma} (-\omega))/(8 \pi^3 G)$.
Similarly, the last parenthesis in  eq. (\ref{Colloperateur1integromega'})
is $H^{\beta\gamma}(-\omega)/8 \pi^3$,
where the matrix $H^{\alpha\beta}(\omega)$ is defined by:
\begin{equation}
H^{\alpha\beta}(\omega) = 8 \pi^3 \! \sum_q m_q^2 \,\sum_{{\mathbf{k}}'} \! \int\!  d^3\!J'\ \ 
\frac{\psi^{\alpha*}_{{\mathbf{k}}'}({\mathbf{J'}})\,  
\psi^{\beta *}_{ - {\mathbf{k}}'}({\mathbf{J'}}) \,
f^q({\mathbf{J'}})  }{
\omega - {\mathbf{k}}'\cdot {\mathbf{\Omega}}({\mathbf{J'}}) }  \ \cdot
\label{defHmatrix}
\end{equation}
The  double Fourier transform $G^{ap}_{ {\mathbf{k}} {\mathbf{k}}'_1}$ of the propagator which is present in the first parentheses
of equations (\ref{Colloperateur1integromega'})~--~(\ref{Colloperateur2integromega'})
can be read from equation (\ref{propagateur1body}):
\begin{equation}
G^{ap}_{ {\mathbf{k}}\, {\mathbf{k}}'_1} (1, 1', \omega) =
\frac{i }{
(\omega - {\mathbf{k}}\cdot {\mathbf{\Omega}}_1)  }
\ \Big(   \frac{1}{8\pi^3}
\delta^a_p \,
\delta({\mathbf{k}}\!\! +\!\! {\mathbf{k}}'_1) \, \delta ({\mathbf{J}}_1\! - \! {\mathbf{J}}'_1)
+\,
\frac{
G m_pm_a \left({\mathbf{k}}\!\cdot\! {\mathbf{\nabla}}_{\mathbf{J}_1} f^a(1)\right)
}{
(\omega + {\mathbf{k}}'_1 \!\cdot\! {\mathbf{\Omega}}'_1) } \ \, 
\psi^\lambda_{\mathbf{k}}({\mathbf{J}}_1) \left(\varepsilon^{-1}(\omega)\right)^{\lambda \mu}
\! \psi^{\mu *}_{-\mathbf{k}'_1}({\mathbf{J}}'_1)
\Big) \ \cdot
\label{propagateurTFTF}
\end{equation}
Using this, the first parenthesis of the second line of equation (\ref{Colloperateur1integromega'}),
$V^\gamma_{a\, 1}(1,\omega)$,  can be written as:
\begin{equation}
V^\gamma_{a\, 1} (1,\omega) = - \frac{i}{8\pi^3}  \, 
\frac{ m_a \left({\mathbf{k}}\cdot\! {\mathbf{\nabla}}_{\mathbf{J}_1} f^a(1)\right)  }{
\omega - {\mathbf{k}}\cdot {\mathbf{\Omega}}_1  }\  \,
\psi^\lambda_{\mathbf{k}}(1)\! \left(\varepsilon^{-1}(\omega)\right)^{\lambda \gamma} \ \cdot
\label{V1}
\end{equation}
The first parenthesis of the second line of 
equation (\ref{Colloperateur2integromega'}), $V^\gamma_{a\, 2} (1,\omega)$,
is similarly calculated
and expressed in terms of the matrix $H$ defined by equation (\ref{defHmatrix}):
\begin{equation}
V^\gamma_{a\, 2} (1,\omega) = \frac{i}{8\pi^3}  \, \frac{ m_af^a(1)  }{ \omega - {\mathbf{k}}\cdot {\mathbf{\Omega}}_1  }
\psi^{\gamma *}_{-\mathbf{k}}(1)  
+ \frac{i}{8\pi^3} \, G m_a \, \frac{\left({\mathbf{k}}\cdot\! {\mathbf{\nabla}}_{\mathbf{J}_1} f^a(1)\right)  }{
\omega - {\mathbf{k}}\cdot {\mathbf{\Omega}}_1  }
\ \, \psi^\lambda_{\mathbf{k}}(1)  \left(\varepsilon^{-1}(\omega)\right)^{\lambda \mu}\! H^{\mu \gamma}(+\omega) \ \cdot
\label{V2}
\end{equation}
Inserting equation (\ref{V1}) in equation (\ref{Colloperateur1integromega'}) we get:
\begin{equation}
{\cal{C}}^{\, a}_1(f) = - \int_B \! \frac{d\omega}{2\pi} \sum_{ {\mathbf{k}}} \
G^2 m_a^2  \, ({\mathbf{k}} \cdot {\mathbf{\nabla}}_{ {\mathbf{J}}_1})  \ \Bigg[ \, 
\psi^\alpha_{-{\mathbf{k}}}(1) \left(\varepsilon^{-1}(-\omega)\right)^{\alpha \beta}
\,  \psi^\lambda_{\mathbf{k}}(1) \left(\varepsilon^{-1}(+\omega)\right)^{\lambda \gamma} 
\! H^{\beta \gamma}(-\omega) 
\ \frac{{\mathbf{k}} \cdot {\mathbf{\nabla}}_{ {\mathbf{J}}_1} f^a(1)}{\omega -  {\mathbf{k}}\cdot {\mathbf{\Omega}}_1  } \, \Bigg] \ \cdot
\label{C1new}
\end{equation}
Inserting equation (\ref{V2}) in equation (\ref{Colloperateur2integromega'}) we get:
$$\displaylines{
{\cal{C}}^{\, a}_2(f) = - \int_B \! \frac{d\omega}{2\pi} \sum_{ {\mathbf{k}}} \
G^2 m_a^2 \,  ({\mathbf{k}} \cdot {\mathbf{\nabla}}_{ {\mathbf{J}}_1}) \ \Bigg[ \,
\psi^\alpha_{-{\mathbf{k}}}(1) \ \, \frac{1}{\omega - {\mathbf{k}}\cdot {\mathbf{\Omega}}_1  } \, \ 
\left((\varepsilon^{-1}(-\omega))^{\alpha\gamma} - \delta^{\alpha\gamma}\right)  
\hfill \cr}$$
\begin{equation}
\qquad \qquad  \Big(\ {1\over G} \, f^a(1) \, \psi^{\gamma *}_{-{\mathbf{k}}}(1) 
+ \left({\mathbf{k}}\cdot\! {\mathbf{\nabla}}_{\mathbf{J}_1} f^a(1)\right) \ \ 
\psi^\lambda_{\mathbf{k}}(1) (\varepsilon^{-1}(+\omega))^{\lambda \mu}  H^{\mu \gamma}(+\omega)\ \Big) \, \Bigg] \ \cdot
\label{C2new}
\end{equation}
Gathering equations (\ref{C1new}) and (\ref{C2new}), the following expression is
obtained for the collision operator:
$$ \displaylines{ 
{\cal{C}}^{\, a}(f) =  - \, \int_B \! \frac{d\omega}{2\pi} \sum_{ {\mathbf{k}}} \
G^2 m_a^2 \, ({\mathbf{k}}\! \cdot\! {\mathbf{\nabla}}_{ {\mathbf{J}}_1})  \ \Bigg[
\, \,  \psi^\alpha_{-{\mathbf{k}}}(1) \
\frac{1}{\omega - {\mathbf{k}}\cdot {\mathbf{\Omega}}_1  }
\hfill \cr } $$
{\vskip -2mm plus .2mm minus .2mm}
$$ \displaylines{
\Bigg[ + (\varepsilon^{-1}(-\omega))^{\alpha \beta} \,  H^{\beta \gamma}(-\omega) \, \psi^\lambda_{\mathbf{k}}(1)
\, (\varepsilon^{-1}(+\omega))^{\lambda \gamma}
\, \left({\mathbf{k}}\cdot\! {\mathbf{\nabla}}_{\mathbf{J}_1} f^a(1)  \right)
\ + \ (\varepsilon^{-1}(-\omega))^{\alpha\gamma} 
\, \psi^\lambda_{\mathbf{k}}(1)\
(\varepsilon^{-1}(+\omega))^{\lambda \mu} H^{\mu \gamma}(+\omega)\, 
\left({\mathbf{k}}\cdot\! {\mathbf{\nabla}}_{\mathbf{J}_1} f^a(1)\right)
\hfill \cr } $$
\begin{equation}
+ \left((\varepsilon^{-1}(-\omega))^{\alpha\gamma} - \delta^{\alpha\gamma}\right)
\psi^{\gamma *}_{-{\mathbf{k}}}(1) {1\over G} \, f^a(1)
\ \,  - \ \, \left({\mathbf{k}}\cdot\! {\mathbf{\nabla}}_{\mathbf{J}_1} f^a(1)\right)
\ \psi^\lambda_{\mathbf{k}}(1)\
(\varepsilon^{-1}(+\omega))^{\lambda \mu} H^{\mu \alpha} (+\omega) 
\ \, \Bigg] \, \Bigg] \ \cdot
\label{Creconstitue}
\end{equation}
The last term on the third line vanishes on integration over $\omega$. Indeed, the Bromwich contour
must pass over all singularities of the function $f(\omega)$ in eq. (\ref{formintegraleomegaprim}), that is,
in equation (\ref{Creconstitue}), over all singularities of functions of $+\omega$.
The contour $B$ can be closed at infinity
in the upper complex plane, which gives, for the fourth term of the square bracket,
a vanishing result. The two terms in the second line of equation (\ref{Creconstitue})
can be associated, yielding
the following expression for ${\cal{C}}^{\, a}(f)$:
$$\displaylines{
{\cal{C}}^{\, a}(f) =  -  \int_B \! \frac{d\omega}{2\pi} \, \sum_{ {\mathbf{k}}_1} \ \, 
G^2 m_a^2  \ ({\mathbf{k}}_1\! \cdot\! {\mathbf{\nabla}}_{ {\mathbf{J}}_1}) \  \Bigg[\,  
\psi^\alpha_{-{\mathbf{k}}_1}(1) \
\frac{1}{\omega - {\mathbf{k}}_1\cdot {\mathbf{\Omega}}_1  } 
\ \, \Big(\left(\varepsilon^{-1}(-\omega))^{\alpha\gamma} - \delta^{\alpha\gamma}\right)
\psi^{\gamma*}_{-{\mathbf{k}}_1}(1) \,\frac{f^a(1)}{G} \Big)\, \Bigg] 
\hfill \cr }$$
\begin{equation}
-  \int_B \!  \frac{d\omega}{2\pi} \,  \sum_{ {\mathbf{k}}_1} \ \, 
G^2 m_a^2 \  ({\mathbf{k}}_1\! \cdot\! {\mathbf{\nabla}}_{ {\mathbf{J}}_1}) \ \Bigg[\, 
\psi^\alpha_{-{\mathbf{k}}_1}(1) \
\frac{\left({\mathbf{k}}_1\cdot\! {\mathbf{\nabla}}_{\mathbf{J}_1} f^a(1)\right)}{\omega - {\mathbf{k}}_1\cdot {\mathbf{\Omega}}_1  }
\ \ (\varepsilon^{-1}(-\omega))^{\alpha\beta} 
\, \psi^\lambda_{\mathbf{k}}(1)\
(\varepsilon^{-1}(+\omega))^{\lambda \gamma} \Big(H^{\beta\gamma}(-\omega)+ H^{\gamma \beta}(+\omega)
\Big)  \, \Bigg]  \cdot
\label{Creconstitue2}
\end{equation}
To evaluate the second line of equation (\ref{Creconstitue2}), the integration contour $B$ may be lowered to the real axis. 
Rigourously, $\omega$ pertains the upper complex half plane and can only be consider real in a limit
sense  when the contour $B$ descends to the real axis. Real singularities at
$\omega = {\mathbf{k}} \cdot {\mathbf{\Omega}}$ must therefore be avoided by the contour by skirting them
from above, so that:
\begin{equation}
\frac{1}{\omega - {\mathbf{k}} \cdot {\mathbf{\Omega}} } \rightarrow
\frac{1}{\omega - {\mathbf{k}} \cdot {\mathbf{\Omega}} + i0} = 
\frac{ {\cal{P}} }{\omega - {\mathbf{k}} \cdot {\mathbf{\Omega}} } - i\pi \delta(\omega - {\mathbf{k}} \cdot {\mathbf{\Omega}} ) \ ,
\end{equation}
where ${\cal{P}}$ is the principle value distribution. Conversely, when $\omega$ descends to
the real axis, $-\omega$ rises to it from below, so that:
\begin{equation}
\frac{1}{-\omega +{\mathbf{k}} \cdot {\mathbf{\Omega}} } \rightarrow
\frac{(- 1)}{\omega - {\mathbf{k}} \cdot {\mathbf{\Omega}} - i0} = 
- \frac{ {\cal{P}} }{\omega - {\mathbf{k}} \cdot {\mathbf{\Omega}} } 
-  i\pi \delta(\omega - {\mathbf{k}} \cdot {\mathbf{\Omega}} ) \ \cdot
\end{equation}
The sum $H^{\beta\gamma}(-\omega)+ H^{\gamma \beta}(+\omega)$ calculated in this limit is:
\begin{equation}
H^{\beta\gamma}(-\omega)+ H^{\gamma \beta}(+\omega) = - 16 i \pi^4 \sum_q \sum_{{\mathbf{k}}'_1 }\int\! d^3\!J'_1 \ 
m_q^2 \ \ \psi^{\gamma *}_{{\mathbf{k}}'_1}(1') 
\psi^{\beta *}_{-{\mathbf{k}}'_1}(1') \, \delta(\omega - {\mathbf{k}}'_1\cdot  {\mathbf{\Omega}}'_1) f^q(1') \ \cdot
\label{HplusH}
\end{equation}
Thanks to the Dirac function in equation (\ref{HplusH}),
the second line of equation (\ref{Creconstitue2}) is easily integrated over $\omega$. 
Where conciseness demands, we note:
$$
\omega_1 = {\mathbf{k}}_1\! \! \cdot\!  {\mathbf{\Omega}}_1 \ , \qquad \qquad \qquad
\omega'_1 ={\mathbf{k}}'_1\! \! \cdot\!  {\mathbf{\Omega}}'_1 \ \cdot
$$
The first line of equation (\ref{Creconstitue2}) can be disposed of by closing the $\omega$
integration contour in the lower half complex plane, which is possible 
because the integrand declines fast enough at infinity. 
The system being supposedly stable,
all the singularities of $(\varepsilon^{-1}(-\omega))^{\alpha\gamma}$ are in the upper half plane. The contour
then encloses only the real singularity at $\omega = {\mathbf{k}}_1 \cdot {\mathbf{\Omega}}_1 $ and its sense  
brings a factor $-2i\pi$  when using the residue theorem. 
The expression of the collision operator then becomes:
$$\displaylines{
{\cal{C}}^{\, a}(f) =  i \ G m_a^2 \ \,  \sum_{{\mathbf{k}}_1 }
\ \, ({\mathbf{k}}_1\! \cdot\! {\mathbf{\nabla}}_{ {\mathbf{J}}_1})
\ \Bigg[ \, 
\psi^\alpha_{- {\mathbf{k}}_1}\! (1) \left((\varepsilon^{-1}(- \omega_1))^{\alpha \beta}
- \delta^{\alpha \beta} \right)  \psi^{\beta *}_{- {\mathbf{k}}_1} (1) f_a(1)
\hfill \cr }$$
\begin{equation}
+ \sum_q \! \sum_{{\mathbf{k}}'_1 } \int\! d^3\!J'_1 \ \, 8\pi^3 G m_q^2 \,
\left(
\psi^\alpha_{- {\mathbf{k}}_1}\! (1)
\left(\varepsilon^{-1}(- \omega'_1)\right)^{\alpha\beta}
\! \psi^{\beta *}_{- {\mathbf{k}}'_1} (1')
\right)
\left( 
\psi^\lambda_{{\mathbf{k}}_1}\! (1)
\left(\varepsilon^{-1}(+\omega'_1)\right)^{\lambda\gamma}
\psi^{\gamma *}_{{\mathbf{k}}'_1} (1')
\right) \
\frac{{\mathbf{k}}_1\!\! \cdot\! \! {\mathbf{\nabla}}_{\mathbf{J}_1} (f^a(1) f^q(1')) }{\omega'_1 - \omega_1  + i\,0}
\ \ \Bigg] \ \cdot
\label{CareconstitAA}
\end{equation}
This expression is then symmetrized. Half of the term on the second line 
of eq. (\ref{CareconstitAA})
is added to half of the same expression, modified by
changing ${\mathbf{k}}_1$ into $-{\mathbf{k}}_1$ and ${\mathbf{k}}'_1$ into $-{\mathbf{k}}'_1$. This  leaves 
it invariant, except for the last denominator, which is changed into
$-(\omega'_1 - \omega_1  - i \, 0)$. 
The half difference brings a contribution 
$- i\pi \delta(\omega'_1 - \omega_1)$.
The term on the first line of equation (\ref{CareconstitAA})
may be similarly symmetrized. When changing ${\mathbf{k}}_1$ to $-{\mathbf{k}}_1$, 
the argument of the inverse response function changes sign. 
The change of the response function when the sign of its real frequency argument, $\omega_r$ say,
is modified may be found by noting that its real and imaginary parts, 
${\varepsilon'}^{\, \alpha\beta}$ and ${\varepsilon"}^{\, \alpha\beta}$, are:
\begin{eqnarray}
&&{\varepsilon'}^{\, \alpha\beta}(\omega_r) = \delta^{\alpha\beta}
- \sum_q \sum_{ {\mathbf{k}}_1 } \!\int\! d^3\! J_1 \ \,  8\pi^3 G m_q^2 
\ \ \psi^{\alpha *}_{{\mathbf{k}}_1}(1) \, \psi^{\beta}_{{\mathbf{k}}_1}(1)
\ {\cal{P}} \left(\frac{ {\mathbf{k}}_1\!\cdot\! {\mathbf{\nabla}}_{\mathbf{J}_1}f^q(1) }{
\omega_r - {\mathbf{k}}_1 \cdot  {\mathbf{\Omega}}_1 }\right) \ ,
\label{epsilonprim}
\\
&&{\varepsilon"}^{\, \alpha\beta}(\omega_r) = \sum_q \sum_{{\mathbf{k}}_1} \!\int\! d^3\! J_1 \, \ 8 \pi^4  G m_q^2 
\ \ \psi^{\alpha *}_{{\mathbf{k}}_1}(1) \,  \psi^{\beta}_{{\mathbf{k}}_1}(1)
\ \left({\mathbf{k}}_1\!\cdot\! {\mathbf{\nabla}}_{\mathbf{J}_1}f^q(1)\right) 
\, \delta( \omega_r - {\mathbf{k}}_1\cdot  {\mathbf{\Omega}}_1) \ \cdot
\label{epsilonseconde}
\end{eqnarray}
The conjugation relations (\ref{psipsichapeau}) can be used to show that: 
\begin{equation}
\varepsilon^{\alpha \beta}(-\omega_r) =  ( \varepsilon^{ {\hat{\alpha}} {\hat{\beta}} }(+ \omega_r))^*
\ , \qquad \qquad \qquad 
( \varepsilon^{-1}(-\omega_r) )^{\alpha \beta}  = 
( (\varepsilon^{-1}(+\omega_r))^{ {\hat{\alpha}} {\hat{\beta}} })^* \ ,
\end{equation}
where the basis element ${\hat{\alpha}}$ associated with $\alpha$ is defined by equation (\ref{hatetstar}).
Using this, the expression $T$, defined by:
\begin{equation}
T \equiv \sum_{ {\mathbf{k}}_1} \  ({\mathbf{k}}_1 \cdot\! {\mathbf{\nabla}}_{ {\mathbf{J}}_1}) 
\ \Bigg[\, 
\psi^\alpha_{- {\mathbf{k}}_1}\! (1) 
\left( 
(\varepsilon^{-1}(- {\mathbf{k}}_1\!\! \cdot\! {\mathbf{\Omega}}_1))^{\alpha \beta}
\!- \!\delta^{\alpha \beta} \right) 
\psi^{\beta *}_{- {\mathbf{k}}_1}\! (1) \, f^a(1)\, \Bigg] \ ,
\label{definitionT}
\end{equation}
which is present in the first line of equation (\ref{CareconstitAA}) is symmetrized to:
\begin{equation}
T  = \, - \, \sum_{ {\mathbf{k}}_1} \ {1\over 2} \,
({\mathbf{k}}_1\! \cdot\! {\mathbf{\nabla}}_{ {\mathbf{J}}_1})\, 
\psi^\alpha_{\mathbf{k}_1}\! (1) \ 
\Big(\! \left(\varepsilon^{-1}(\omega_1)\right)^{\alpha\beta}
\! - \left(\left(\varepsilon^{-1}(\omega_1)\right)^{\beta \alpha} \right)^* \Big)  
\ \psi^{\beta *}_{\mathbf{k}_1} (1) \, f^a(1) \ \cdot
\label{TTT}
\end{equation}
As expected \citep{NelsonTremaine}, this expression involves the antihermitian 
part of the matrix $\varepsilon^{-1}$, which may be expressed
in terms of the antihermitian part of the matrix $\varepsilon$ as:
\begin{equation}
 \left(\varepsilon^{-1}\right) -  \left(\varepsilon^{-1}\right)^\dagger =
\varepsilon^{-1} (\varepsilon^\dagger - \varepsilon) \, (\varepsilon^{\dagger})^{-1} \ \cdot
\end{equation}
The matrix $\varepsilon^\dagger -\varepsilon$ is calculated from equations 
(\ref{epsilonprim})~--~(\ref{epsilonseconde}):
\begin{equation}
(\varepsilon^{\beta\alpha}(\omega))^* - \varepsilon^{\alpha\beta}(\omega) = 
-i \sum_q \sum_{ {\mathbf{k}}'_1} \ \, 16\pi^4 G m_q^2   \int\! d^3\! J'_1
\ \, \delta(\omega - {\mathbf{k}}'_1\! \cdot\! {\mathbf{\Omega}}'_1)
\ \left({\mathbf{k}}'_1 \! \cdot \! {\mathbf{\nabla}}_{ { {\mathbf{J}}' }_1 } f^q(1') \right)
\psi^{\alpha *}_{{\mathbf{k}}'_1}(1') \, \psi^{\beta}_{{\mathbf{k}}'_1}(1') \ \cdot
\end{equation}
The term $T$ in equation (\ref{TTT}) can then be written as:
\begin{equation}
T \! =\!  8 i\pi^4 \sum_q \sum_{ {\mathbf{k}}_1} \! \sum_{ {\mathbf{k}}'_1} \!\int\! d^3\!J'_1
\ \ G m_q^2 \ \delta({\mathbf{k}}_1\!\cdot  {\mathbf{\Omega}}_1 - {\mathbf{k}}'_1\!\cdot  {\mathbf{\Omega}}'_1)
\  ({\mathbf{k}}_1\! \cdot {\mathbf{\nabla}}_{ {\mathbf{J}}_1})
\ \Bigg[\, \left\vert\, 
\psi^\lambda_{ {\mathbf{k}}_1}(1) (\varepsilon^{-1}(\omega_1))^{\lambda \mu}
\psi^{\mu *}_{ {\mathbf{k}}'_1}(1')\, \right\vert^2 
({\mathbf{k}}'_1\! \cdot\! {\mathbf{\nabla}}_{ {\mathbf{J}}'_1} f^q(1')) \, \Bigg] 
\label{Tlast}
\end{equation}
and inserted in the first line of equation (\ref{CareconstitAA}).
The square modulus factor in  equation (\ref{Tlast}) is 
$\mid\!{\cal{D}}_{{\mathbf{k}}_1 {\mathbf{k}}'_1}({\mathbf{J}}_1, {\mathbf{J}}'_1, \omega_1)\!\mid^{-2}$
(equation (\ref{definitionD})).
The second line of equation (\ref{CareconstitAA}) can be treated similarly. From equation (\ref{definitionD}),
one of the parentheses is 
$( {\cal{D}}_{{\mathbf{k}}_1 {\mathbf{k}}'_1}({\mathbf{J}}_1, {\mathbf{J}}'_1, \omega'_1))^{-1}$ 
and the other is its complex conjugate, which can be shown by using 
the conjugation relation (\ref{psipsichapeau}). 
When all these symmetrizations  and substitutions are made, 
the collision operator is finally written as:
\begin{equation}
{\cal{C}}^{\, a}(f) =
\sum_q \sum_{ {\mathbf{k}}_1 } \sum_{ {\mathbf{k}}'_1 }\! \int\! d^3\! J'_1  
\ \, 8\pi^4 G^2 m_a^2 m_q^2 \  \, 
{\mathbf{k}}_1\! \cdot\! {\mathbf{\nabla}}_{ {\mathbf{J}}_1} \ \left[\, 
\frac{\delta({\mathbf{k}}_1\!\cdot\!  {\mathbf{\Omega}}_1 - {\mathbf{k}}'_1\!\cdot\!  {\mathbf{\Omega}}'_1) }{
\left\vert 
{\cal{D}}_{{\mathbf{k}}_1 {\mathbf{k}}'_1}({\mathbf{J}}_1, {\mathbf{J}}'_1, {\mathbf{k}}_1\!\cdot\!  {\mathbf{\Omega}}_1)
\right\vert^2  }
\, \left({\mathbf{k}}_1\! \cdot\! {\mathbf{\nabla}}_{ {\mathbf{J}}_1} \!
- \! {\mathbf{k}}'_1\! \cdot\! {\mathbf{\nabla}}_{ {\mathbf{J}}'_1 }\right) f^a({\mathbf{J}}_1)
f^q({\mathbf{J}}'_1) \, \right] \ \cdot
\label{lequationapp}
\end{equation}

\section{Variables and Fourier coefficients for spherical potentials}
\label{sphericalities}

\subsection{Angle and action variables for a spherically symmetric potential}
\label{secspherical}

The motion of a particle  in a spherically symmetric potential is best described in
spherical coordinates $r$, $\theta$, $\varphi$, the variable $r$ being the distance to
the center, $\theta$ the colatitude measured from the pole associated with
the coordinate polar axis $z$ and $\varphi$ the azimuth measured from an arbitrarily defined
origin in the equatorial plane. Let $U(r)$ be the gravitational potential, an increasing function of $r$
approaching 0 at infinity, which
is provisionnally treated as constant in time. The fact that $U(r)$ actually slowly evolves
as the relaxation proceeds is addressed in section \ref{secsystemcouple}.
Without loss of generality, the particle may be assumed to be of unit mass.
A dot indicating time derivative,
the conjugate momenta to $r$, $\theta$, $\varphi$ are:
\begin{equation}
p_r = {\dot{r}} \qquad p_\theta = r^2 {\dot{\theta}} \qquad p_\varphi = r^2 \sin^2 \theta \  {\dot{\varphi}} \ \cdot
\end{equation}
In a constant potential, the energy $E$ of a particle
is a first integral, as is the vectorial angular momentum
${\mathbf{L}}$, that is, its modulus $L$,
its projection $L_z$ on the polar axis and the direction of its projection onto the equatorial plane.
The angle and action variables are deduced from the variables $r$, $\theta$, $\varphi$, $p_r$, $p_\theta$, $p_\varphi$
by a canonical transformation, the generating function of which is the solution to the Hamilton-Jacobi equation.
\citet{Goldstein} shows how to construct
angle and action variables $w_1$, $w_2$, $w_3$, $J_1$, $J_2$, $J_3$
in the case of a newtonian potential. Similar results for a general spherical potential
are also well known. They can be found, for example,  in \citet{WeinbergTremaine}  or in \citet{Saha}.
A summary is presented in this appendix.

\bigskip

The orbit in a spherically symmetric potential being plane,
the periods of the azimuthal and latitudinal motions are equal. This introduces some freedom  in defining the
actions, which can be taken advantage of to impose that one of the angles, $w_3$ say, be a first integral, associated with
the direction of the equatorial projection of the angular momentum.
The origin of the constant angle $w_3$ can be chosen to coincide with the origin of the azimuths and the origin of the
radial angle variable $w_1$ may be placed at some fiducial periapse.
The angle and action variables then are given by the following expressions:
\begin{eqnarray}
&&J_1 = \frac{1}{\pi} \int_{r_P}^{r_A} \sqrt{ 2(E - U(r)) - J_2^2/r^2} \ \, dr\, ,
\qquad \qquad J_2 = L = (p_\theta^2 + p_\varphi^2/\sin^2 \theta)^{1/2},
\qquad \qquad J_3 = L_z = p_\varphi\, ,
\label{Jspheriq} \\
&&w_1 = \pm \! \int_{P}^{M}\! \frac{\Omega_1 \mid\!dr'\!\mid}{\sqrt{ 2(E - U(r')) - J_2^2/r'^2}}\, ,
\qquad  w_2 = \psi \pm \! \int_{P}^{M}\! \frac{(\Omega_2 - L/r'^2) \mid\!dr'\!\mid}{\sqrt{ 2(E - U(r')) - J_2^2/r'^2}}\, ,
\quad  w_3 = \varphi - \arcsin \left(\cot \theta \cot \beta\right) \cdot
\label{wspheriq}
\end{eqnarray}
$P$ represents the position
of a particle passing at the point $M = r,\theta, \varphi$, with momenta $p_r$, $p_\theta$, $p_\varphi$
when it reaches a fiducial periapse of its orbit.
\begin{figure}
\begin{center}
\includegraphics[scale=0.7]{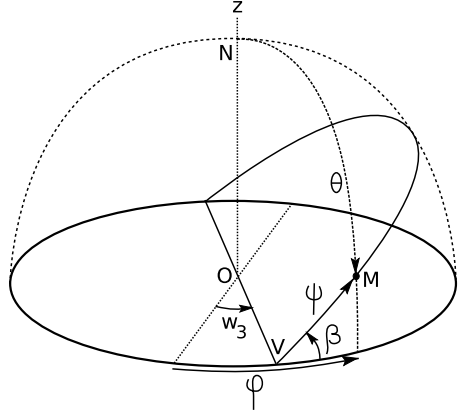}
\end{center}
\caption{\small Angular parameters associated with the projection M of the
particle on the unit sphere. O is the origin of the spherical coordinates,  N the origin of
the colatitudes and V the ascending node of the orbit. The orbital plane is OVM.}
\label{figangles}
\end{figure}
For given position and momenta, the action and angle variables in equations
(\ref{Jspheriq})--(\ref{wspheriq}) depend on the potential $U(r)$.
The radii $r_P$ and $r_A$ are the distances to the origin of the periapses and the apoapses 
of the orbit of a particle with actions ${\mathbf{J}} = (J_1,\, J_2,\, J_3)$. They are
given by the equation:
\begin{equation}
2(E - U(r)) - J_2^2/r^2 = 0 \ ,
\label{periapoequation}
\end{equation}
and they depend on $E$ and $J_2$, that is, on $J_1$ and $J_2$ but not on $J_3$.
There being many different periapses, the fiducial one must be defined not only
by its spatial location, but also by the time at which the particle passes there.
The sign $\pm$ in equations (\ref{wspheriq})
should be taken as $+$ when the particle visits the fiducial periapse $P$ before it passes at $M$
and $-$ otherwise. $\Omega_1$ and $\Omega_2$ are the pulsations  of the radial and latitudinal motions respectively
(equations (\ref{Omegaspheriq})).
The angle  $\psi$ is the azimuth of the present particle's position in the orbital plane,
measured from the ascending node (figure \ref{figangles}). The angle $w_2$, which varies linearly
in time, is the mean angular motion of the particle in the plane of its orbit.
The constant angle $w_3$ is the azimuth of the ascending node in the equatorial plane.
The angles $w_1$ and $w_2$ are expressed as radial integrals
following the sense of the particle's motion, whence
the presence of an absolute value of the differential element
in equations (\ref{wspheriq}). The boundaries of these
integrals on $r'$ have not been written as $r_P$ and $r$ because, depending on the relative position of the particle
and the fiducial periapse, the integral may be extended over several successive senses of the radial oscillations.
The ratio $J_3/J_2$ is the cosine
of an inclination angle $\beta$ (figure \ref{figangles})
defined by $ \cos \beta = J_3/J_2$.
The latitude of the particle oscillates between $\pm \beta$.
The frequency $\Omega_3$ vanishes and the frequencies $\Omega_1$ and $\Omega_2$ are given by:
\begin{equation}
\frac{\pi}{\Omega_1}  = \int_{r_P}^{r_A} \frac{dr}{\sqrt{2 (E - U(r)) - J_2^2/r^2}} \ ,
\qquad  \qquad \qquad \qquad
\Omega_2 = \frac{\Omega_1}{\pi} \  \int_{r_P}^{r_A} \frac{J_2}{r^2} \ \,  \frac{dr}{\sqrt{2 (E - U(r)) - J_2^2/r^2}} \ \cdot
\label{Omegaspheriq}
\end{equation}
They are both positive. The angle variables $w_1$ and $w_2$
then increase linearly with time with the frequencies $\Omega_1$ and $\Omega_2$,
changing by $2\pi$ in a complete, respectively radial and latitudinal, oscillation.
Equations (\ref{Jspheriq})--(\ref{wspheriq}) give the angle and action variables in terms of the
position and momentum variables. These relations may  be inverted to give the latter in terms of the former.
This however involves the inversion of the implicit relation (\ref{Jspheriq}) to obtain $E$ as a function of $J_1$ and  $J_2$
and of the first of equations (\ref{wspheriq}) to obtain $r$ as a function of  $w_1$, $J_1$, $J_2$.

\subsection{Basis Fourier coefficients for spherical potentials}
\label{Psikspheriq}

The basis expansion coefficients which correspond to the density distribution (\ref{densitemassefa})
are obtained from equations (\ref{associatedD}) and (\ref{coeffadephib}).
When, as here, the distribution functions do not depend on angles, their integrals over
angles in equation (\ref{coeffadephib}) are simply proportional to the ${\mathbf{k}} = {\mathbf{0}}$ Fourier coefficient
of $ \psi^{\alpha *}$, or equivalently of $\psi^{\hat{\alpha}}$ (equation (\ref{hatetstar})), which is
the complex conjugate of $\psi^\alpha_{\mathbf{0}}({\mathbf{J}}, t)$
(equation (\ref{psipsichapeau})).
This coefficient depends on time, due to the slow variation of the orbits. We find that:
\begin{equation}
a_\alpha (t) = - \! \sum_a \,  8 \pi^3 m_a \int\!\! d^3\! J 
\ \ f^a({\mathbf{J}},t) \ (\psi^{\alpha}_{\mathbf{0}}({\mathbf{J}}, t))^* \ \cdot
\end{equation}
From equations (\ref{DeltaetPsi}) and (\ref{phiexpansiongene}),
the gravitational potential is $U({\mathbf{r}}, t) =  G\, a_\alpha(t) \, \psi^{\alpha}({\mathbf{r}})$,
$\alpha$ being a dummy index. Its partial time derivative is given by equation (\ref{variationpot}).
We also need some explicit expression for the angle Fourier coefficients $\psi^{\alpha}_{{\mathbf{k}} }({\mathbf{J}}, t)$
We also need some explicit expression for the angle Fourier coefficients $\psi^{\alpha}_{{\mathbf{k}} }({\mathbf{J}}, t)$
of the basis potentials (equation (\ref{LATFdepsi})).
The relation of the position ${\mathbf{r}}$ to the angle and action variables ${\mathbf{w}}$ and ${\mathbf{J}}$
depends on the potential $U({\mathbf{r}},t)$, and thus on $t$.  One could think of evaluating
$\psi^{\alpha}_{{\mathbf{k}} }({\mathbf{J}}, t)$ for a given potential $U({\mathbf{r}}, t)$
by just calculating the integral over angles in equation (\ref{LATFdepsi}). The position vector
${\mathbf{r}}$ would then have to be expressed in terms of the angle vector ${\mathbf{w}}$, for given actions. This
cannot be done explicitly in general, since
the relations (\ref{Jspheriq})--(\ref{wspheriq}) would have to be inverted.
For spherical potentials, it is easier to change the variables of integration
$w_1$, $w_2$, $w_3$ in equation (\ref{LATFdepsi}) for position-type variables $r$, $\psi$, $w_3$
(figure \ref{figangles}). For given actions ${\mathbf{J}}$, these variables
are related by the equations (\ref{wspheriq}) which can also be written,
with the notations explained 
above, as $w_1 = W_1({\mathbf{J}}, M(r), t)$ and
$w_2 = \psi + W_2({\mathbf{J}}, M(r), t)$, where:
\begin{equation}
W_1({\mathbf{J}}, M(r), t) =\!\! \int_{P}^{M(r)}\!\!\!
\frac{\Omega_1(t) \, \mid \! dr'\!\mid}{\sqrt{2(E(t) - U(r',t)) - J_2^2/r^{'2} }}
\ , \qquad  \qquad
W_2({\mathbf{J}}, M(r), t) =
\!\! \int_{P}^{M(r)}\!\!\!
\frac{(\Omega_2(t) - {J_2^2}/{r^{'2}}) \mid \! dr'\!\mid}{\sqrt{2(E(t) - U(r',t)) - {J_2^2}/{r^{'2}} }} \ \cdot
\label{fonctionW}
\end{equation}
The jacobian of the transformation from $w_1$, $w_2$, $w_3$ to $r$, $\psi$, $w_3$ is $\mid\!dW_1/dr\!\mid$.
For a  spherical potential, the basis potential functions can be chosen to depend only on
the radial distance $r$.
Equation (\ref{LATFdepsi}) then becomes:
\begin{equation}
\psi^{\alpha}_{{\mathbf{k}} }({\mathbf{J}},t) =  \oint \! \mid \! dr\!\mid
\!\!\int_0^{2\pi}\! d\psi \! \int_0^{2\pi}\! dw_3 \,
\frac{\Omega_1(t) \, \psi^\alpha(r)
}{\sqrt{2 (E - U(r,t)) - {J_2^2}/{r^2}}}
\  \exp(-i (k_1 W_1(M(r),t) + k_2 W_2(M(r),t) + k_2 \psi + k_3 w_3)) \ \cdot
\label{psikspheriqvarposition}
\end{equation}
The integrations over the
angles $\psi$ and $w_3$ reduce to $4\pi^2 \delta_{k_2}^0 \, \delta_{k_3}^0$  where the $\delta$'s are Kronecker symbols.
Thus the coefficients $\psi^{\alpha}_{{\mathbf{k}} }$ differ from zero
only when the $k_2$ and $k_3$ components vanish.
The radial integration is over a complete oscillatory cycle of the variable $r$, from $r_P$ to $r_A$ and back.
The coefficients $\psi^{\alpha}_{k}({\mathbf{J}},t)$ depend on
the potential $U(r,t)$ and on the $k_1$ component of $\mathbf{k}$, hereafter simply noted $k$.
Equation (\ref{psikspheriqvarposition}) then reduces to:
\begin{equation}
\psi^{\alpha}_{k}({\mathbf{J}}) = 4\pi^2 \Omega_1(t) \,
\oint \! \mid \! dr\!\mid
\frac{\psi^\alpha(r) \ e^{-i k W_1(M(r),t)} }{\sqrt{2 (E - U(r,t)) - J_2^2/r^2}} \ \cdot
\label{psiradial}
\end{equation}
The cycle integral over $r$ in equation (\ref{psiradial})
can be separated into an ascending part, in which $r$ increases from $r_P$ to $r_A$, and a descending part in
which it decreases from  $r_A$ to $r_P$. Let $W_1^+(r)$
be the value of $W_1(M(r), t)$ during the ascending part
and $W_1^-(r)$
its value during the descending part.
$W_1(M(r),t)$ is a monotonically increasing function along the oscillation. Its value at the apoapse is $\pi$.
Equation (\ref{fonctionW}) shows that $\pi - W_1^+(r)= W_1^-(r) - \pi$.
Defining $W_{k}(r,t) = k W_1^+(r,t)$, equation (\ref{psiradial}) is turned into equation (\ref{W+k1k2}).

\bigskip

\bigskip

{\bf{Accepted by MNRAS,  2010, April 22~;  Received 2010 April 13~; in original form 2010 February 25}}
\end{appendix}

\end{document}